\documentclass[letterpaper, 12pt]{article} 
\usepackage[margin=1in]{geometry}
\newcounter{somecounter}
\usepackage{booktabs, array}
\usepackage{setspace}
\usepackage{float}
\usepackage{epstopdf}
\usepackage{standalone}
\usepackage{color}
\usepackage{xcolor}
\usepackage{bm}
\usepackage{amsmath,amsfonts,amssymb}
\usepackage{amsthm}

\usepackage{pdflscape}
\usepackage{authblk}
\usepackage{graphicx}
\usepackage[colorlinks=true,citecolor=black,linkcolor=black,hidelinks]{hyperref}
\usepackage[normalem]{ulem}
\usepackage{multirow}
\usepackage{multicol}
\usepackage[flushleft]{threeparttable}
\usepackage{enumitem}
\usepackage{url}
\usepackage[utf8]{inputenc}
\usepackage{amsmath}
\usepackage{amsfonts}
\usepackage{amsthm}

\usepackage{lipsum}

\usepackage{natbib}
\usepackage{mathtools}

\newcommand{\interior}[1]{%
 {\kern0pt#1}^{\mathrm{o}}%
}
\usepackage[english]{babel}
\usepackage[autostyle, english = american]{csquotes}
\MakeOuterQuote{"}

\usepackage{algorithm,algpseudocode}
\usepackage{caption}
\usepackage{subcaption}
\usepackage[makeroom]{cancel}

\makeatletter
\newcommand*\bigcdot{\mathpalette\bigcdot@{.5}}
\newcommand*\bigcdot@[2]{\mathbin{\vcenter{\hbox{\scalebox{#2}{$\m@th#1\bullet$}}}}}
\makeatother

\title{\textbf{Scalable, Likelihood-Free Calibration of Ice-Sheet Models with Deep Diffusion Emulators and Feature Matching}}

\author[1]{Kanghyun Wi}
\author[1,2]{Jaewoo Park}
\author[3]{Saumya Bhatnagar}
\author[4,5,*]{Won Chang}

\affil[1]{Department of Statistics and Data Science, Yonsei University, Seoul, Republic of Korea}
\affil[2]{Department of Applied Statistics, Yonsei University, Seoul, Republic of Korea}
\affil[3]{Crop Science Division, Bayer AG, Fremont, CA, USA}
\affil[4]{Department of Statistics, Seoul National University, Seoul, Republic of Korea}
\affil[5]{Institute for Data Innovation in Science, Seoul National University, Seoul, Republic of Korea}
\affil[*]{Corresponding author. Email: \href{Email:wonchang@snu.ac.kr}{wonchang@snu.ac.kr}}

\begin{document}

\def\spacingset#1{\renewcommand{\baselinestretch}%
{#1}\small\normalsize} \spacingset{1}

\maketitle

\begin{abstract}
The Antarctic ice sheet is a major source of uncertainty in future sea-level projections, and physical simulators such as the PSU3D-ICE model are essential for studying its evolution. Calibrating them against observations is challenging: the simulator outputs and observed ice-thickness fields are high-dimensional, spatially dependent, and semi-continuous, with a large point mass at zero denoting ice-free regions. These features make conventional Gaussian-process emulation and likelihood-based calibration ill-suited and computationally infeasible at full resolution. We propose the sequential calibration method guided by a diffusion model and a Siamese network (SC-DS), a fully neural framework. For emulation, we develop a single conditional diffusion model that jointly generates the binary ice presence--absence pattern and the continuous thickness field, using global and local conditioning to represent how the input parameters shape the output. Because the emulator induces an intractable likelihood, our likelihood-free calibration replaces the hand-chosen distance and tolerance of approximate Bayesian computation with a probabilistic acceptance rule learned by an iteratively retrained Siamese network, together with a data--model discrepancy adjustment. Applied to the West Antarctic Ice Sheet, SC-DS matches the accuracy of state-of-the-art Gaussian-process calibration at a fraction of its computational cost and scales to the full-resolution domain, where existing methods become intractable.
\end{abstract}

\noindent%
{\it Keywords:} Computer model calibration, computer model emulation, diffusion model, semi-continuous spatial data, Antarctic ice sheet 

\spacingset{1.8} 

\section{Introduction}
\label{sec: Introduction}
The impacts of climate change exert a profound influence on the cryosphere. In particular, the West Antarctic Ice Sheet (WAIS) is one of the most climate-sensitive components of the cryosphere, primarily because most of its mass lies below sea level. It is significantly affected by temperature fluctuations, leading to processes such as basal melting, dynamic retreat, and substantial mass loss. These processes directly translate the signal of climate change into substantial contributions to global sea-level rise \citep{deschamps2012, bedmap3}. Given that rising sea levels pose direct risks to human settlements and critical infrastructure, cryospheric changes have important societal implications. A substantial proportion of the global population resides in regions near the current sea level. Accordingly, the scientific urgency of understanding the governing dynamics of the WAIS is amplified by the need to provide credible assessments of future sea-level hazards to vulnerable coastal populations. Consequently, understanding and inferring WAIS evolution is crucial for managing global climate risk.

Simulation using computer models plays a key role in understanding and projecting the future dynamics of the WAIS \citep{PollardandDeConto2012}. However, there is substantial uncertainty in the specification of key model parameters that represent a multitude of glaciological processes, such as ice-bedrock and ice-ocean interactions and subglacial hydrology. Because it is not possible to directly observe these phenomena, indirect sources of information, such as the spatial pattern of ice thickness in the WAIS, are used to infer these parameters. This, however, poses its own statistical challenges because the observed and modeled ice-thickness patterns take the form of high-dimensional, spatially correlated, and zero-inflated data \citep{StonEtAl2010, Chang2022}. Therefore, developing a calibration framework capable of accommodating such complex data would be a crucial step toward addressing these scientific challenges.

As noted above, calibrating ice-sheet models presents a substantial challenge not only because the underlying data are spatially correlated and high-dimensional, but also because they exhibit a semi-continuous distribution driven by an excess of zeros. More specifically, both the computer model outputs and the observational ice-thickness fields take positive values in regions where ice is present and zero values where it is absent. This high-dimensional zero-inflated structure renders the seminal calibration framework proposed by \cite{KennedyandHagun2002} infeasible. Furthermore, most existing studies of the Antarctic ice sheet have relied on partial or highly aggregated data representations \citep{Chang2022, Lee2020}. Although their research demonstrated meaningful parameter uncertainty quantification, it is reasonable to expect that restricting the analysis to a subset of available data may lead to a loss of information. This caveat motivates a calibration approach that uses the full high-dimensional ice patterns.

Another issue is the high computational cost of ice models. As with other physical models that produce high-dimensional outputs, each run of an ice model incurs a substantial computational burden; thus, we can access only a limited number of model outputs. The standard surrogate approach for such problems is Gaussian process (GP) emulation. Although numerous studies have achieved meaningful inferences using this technique \citep[see, e.g.,][]{KennedyandHagun2002, Higdon2008, Mengyang2016, Chang2022}, applying GPs to high-dimensional data can introduce its own computational burden. This is because the GP-based approach involves inverting a matrix, which is computationally expensive when the data are high-dimensional. Furthermore, the GP-based approach is challenging for high-dimensional data with an excess of zeros, owing to the inherent nature of Gaussian variables. \cite{Chang2022} provided a solution based on a mixture model and dimension reduction, but the approach scales only to data with thousands of spatial locations. Consequently, \cite{Chang2022} used only a small portion of the WAIS spatial pattern, potentially losing information. Therefore, analyzing full-scale WAIS ice patterns requires emulation methods that efficiently handle semi-continuous, high-dimensional spatial data.

In this article, we propose the sequential calibration method guided by diffusion models and a Siamese network (SC-DS), a neural network-based framework for emulation and likelihood-free calibration. We use diffusion models \citep{ HoEtAl2020}, a state-of-the-art class of deep generative models, to emulate the computer model. Compared with GP-based emulators, diffusion models efficiently handle high-dimensional data via GPU computation. Nevertheless, since the standard diffusion model also relies on Gaussian distributions, it faces challenges in adequately handling semi-continuous data with an excess of zeros. To address this, we employ a single conditional diffusion model that jointly generates the binary ice presence-absence pattern \citep[via the Bit diffusion representation][]{Chen2022AnalogBG} and the continuous thickness pattern, yielding accurate semi-continuous model outputs

However, the use of a diffusion model-based emulator poses a challenge in the calibration stage. Existing calibration approaches \citep{KennedyandHagun2002} are not directly applicable here because the implicit likelihood of the diffusion model is intractable \citep{Li2023}. Although approximate Bayesian computation (ABC) is one of the most standard alternatives for circumventing this problem \citep{Marinabc}, it can still be computationally inefficient when the data are high-dimensional or the parameters are highly uncertain. To address this issue, SC-DS employs a Siamese network \citep{BromleyEtAl1993} to construct a likelihood-free calibration rule. 

The Siamese network consists of two weight-sharing convolutional neural networks (CNNs) that project high-dimensional input fields into low-dimensional feature representations, whose comparison provides a data-driven measure of similarity between the target observation and candidate model outputs. Using this similarity measure, we identify input parameters that generate outputs most similar to the target observation without evaluating the likelihood. We also employ a sequential approach that propagates the proposal distribution through a sequence of intermediate distributions, thereby enabling more efficient exploration of the parameter space. As this procedure makes the proposal-generated outputs increasingly similar to the target observation, the discrimination task for the Siamese network becomes more difficult; we therefore retrain the network iteratively to maintain an informative similarity measure throughout the calibration procedure.

An alternative is neural posterior estimation (NPE) \citep{Papamakarios2016, greenberg2019}, which directly models the posterior distribution. However, designing an NPE architecture capable of accommodating the high-dimensional, zero-inflated nature of our spatial data remains highly challenging. Moreover, the large number of simulations required to train such models is computationally prohibitive for our expensive computer model. These limitations motivate our proposed approach. By combining an emulator with an iteratively retrained Siamese network, our framework avoids the cost of repeatedly running the computer model while automatically extracting informative summaries from the complex spatial data.

The remainder of this paper is organized as follows. Section~\ref{sec:computer-model-and-data-description} describes the PSU3D-ICE model, the ensemble of model runs, and the Bedmap3 observational data used throughout the study. Section~\ref{sec: Emulation Using the Diffusion Model} develops the diffusion-based emulator, detailing how the binary ice presence-absence pattern and the continuous thickness field are jointly generated within a single conditional architecture and how the input parameters are injected through global and local conditioning mechanisms. Section~\ref{sec: Likelihood-Free Calibration} presents the likelihood-free calibration procedure, including the Siamese-network acceptance rule, the sequential sampling scheme with iterative retraining, and the data--model discrepancy adjustment. Section~\ref{sec: Applications} evaluates the proposed framework: we first benchmark it against the state-of-the-art Gaussian-process method of \citet{Chang2022} on a reduced spatial domain, then demonstrate that it scales to the full-resolution WAIS through emulator diagnostics and calibration on both a held-out model run and the real Bedmap3 observations. Section~\ref{sec: Discussion} concludes with a summary of our findings and directions for future work.

\section{Computer Model and Data Description}
\label{sec:computer-model-and-data-description}
In this article, we use a widely used Antarctic ice model, the Pennsylvania State University 3D ice sheet model (PSU3D-ICE model) \citep{PollardandDeConto2012, sadai2025antarctic, halberstadt2026antarctic}, to study the behavior of the WAIS. This model enables simulation of the long-term evolution of the WAIS at high spatial resolution (10--40 km). As in other studies aimed at calibrating complex computer models, running the PSU3D-ICE model within a calibration procedure is computationally infeasible: each run takes about 30 minutes, and a typical calibration procedure requires running the model at least thousands or tens of thousands of times. We therefore adopt an emulation approach that evaluates a limited number of prespecified runs at selected design points and builds a fast surrogate model from them.

We use 499 model runs from the PSU3D-ICE model, with parameter settings selected using Latin hypercube sampling. (We originally intended to generate 500 runs, but one failed due to numerical instability.) The simulation results comprise the full ice-thickness pattern, including both floating ice shelves over the ocean and grounded ice, the portion of the ice sheet resting on bedrock.

The calibration approach proposed in this study aims to infer 10 selected input parameters, each with a physically interpretable meaning in the PSU3D-ICE model. Although these parameters have different ranges \citep{PollardandDeConto2012, PollardEtAl2015}, they are all rescaled to the interval $[0,1]$ for easier computation. These parameters are highly uncertain and therefore difficult to calibrate; however, they are known to exert strong control over the modern WAIS configuration. Consequently, inferring these unknown parameters from observational data represents a problem of both geophysical and statistical significance. A detailed description of the input parameters is provided in Section~A of the Supplementary Material.

The model starts from the last glacial maximum and runs for 40,000 years to reach the present. Atmospheric forcing uses the modern Antarctic climate dataset \citep[ALBMAP;][]{LeBrocqEtAl2010}, with uniform cooling adjustments scaled to a deep-sea $\delta^{18}\mathrm{O}$ record \citep{PollardDavid2009, PollardandDeConto2012}. Oceanic forcing is based on archived ocean temperature fields from a coupled AOGCM simulation covering the last 20,000 years \citep{Liu2009}. 

\begin{figure}[ht]
    \centering
    \includegraphics[width=0.8\textwidth]{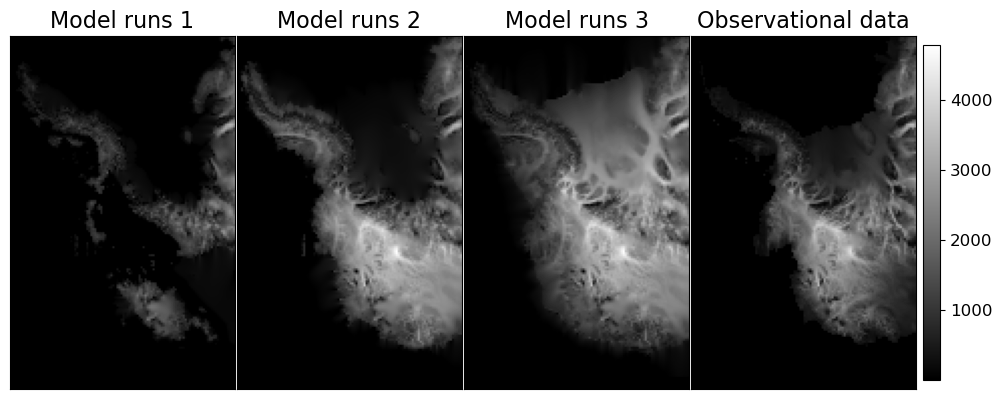}
    \includegraphics[width=0.9\textwidth]{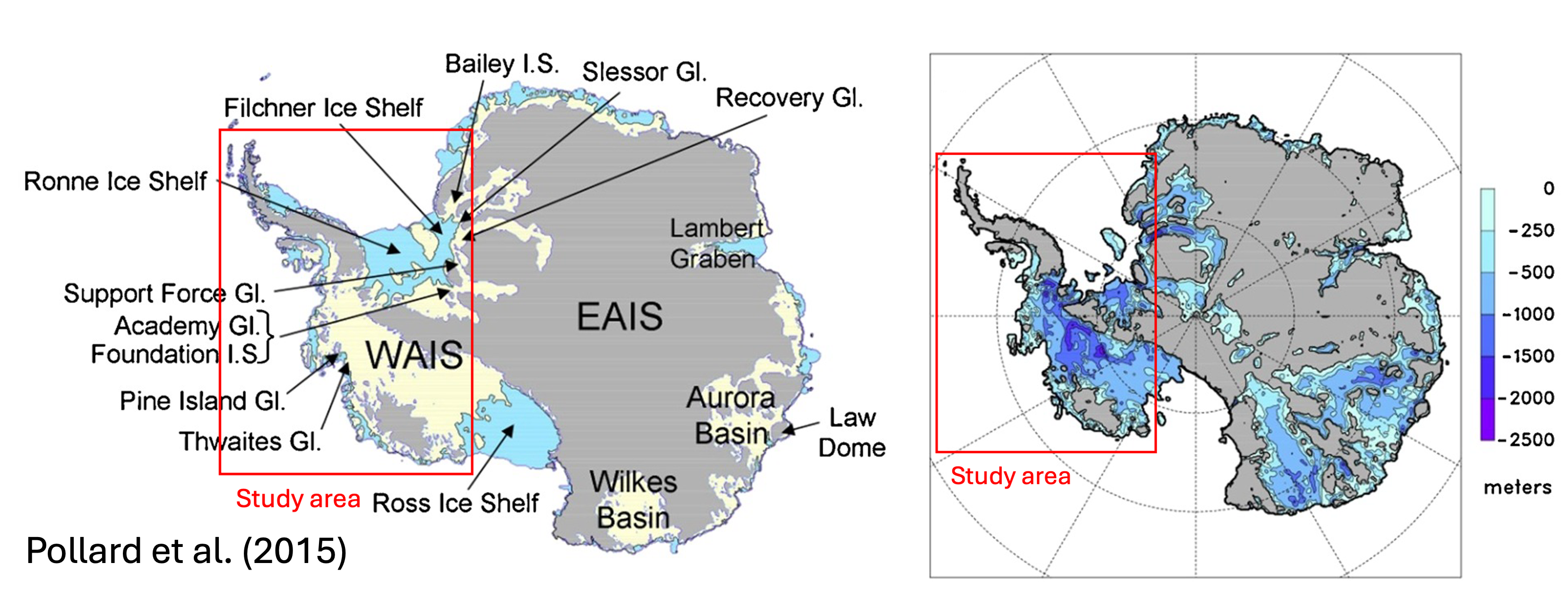}
    \caption{From top to bottom, left to right. Top: PSU3D-ICE model outputs with varying degrees of ice coverage, together with the observational data from the Bedmap3 dataset \citep{bedmap3}. Bottom: a map with major geographic labels (left) and a modern bedrock elevation map (right), adapted from \citet{PollardEtAl2015}. The study area is highlighted by a rectangle.}
    \label{fig: Examples and observational data}
\end{figure}

After simulating the full WAIS domain, we extract the modern ice-thickness pattern. The designated domain is a $176 \times 112$ grid with a 20 km resolution in $\mathbb{R}^2$. Representative examples of the simulation runs, the observational ice-thickness data, and the study area are shown in Figure~\ref{fig: Examples and observational data}. To calibrate the ten input parameters, we compare model outputs with the observed ice thickness over the same region. Since the observational data set has a higher resolution (1 km), we rescale it to match the model output resolution by averaging over 400 pixels per model pixel (reflecting the 20-fold difference in resolution, i.e., $20^2 = 400$). Note that both the model outputs and the observational data are high-dimensional semi-continuous spatial data: the distribution has a positive probability mass at zero, indicating the absence of ice, and a continuous distribution over positive values, denoting the thickness of the ice sheet at each location. This non-Gaussian distribution introduces significant statistical challenges for emulation and calibration.

\section{Emulation Using the Diffusion Model}
\label{sec: Emulation Using the Diffusion Model}
In this section, we describe the emulation component of SC-DS, which uses a diffusion model to emulate the ice model output. The overall model architectures are illustrated in Section~B of the Supplementary Material.
Let $Y(\boldsymbol{\theta},\mathbf{s})$ be a model output at grid point $\mathbf{s}\in \mathcal{S}$ with input parameter $\boldsymbol{\theta} \in \boldsymbol{\Theta}$ where $\mathcal{S}\subset \mathbb Z^2$ is the spatial region that covers the area of interest and $\boldsymbol{\Theta}\subset \mathbb R^{10}$ is the parameter space. Let $\mathbf{Y}(\boldsymbol{\theta})$ be an $n$-dimensional vectorized model output with parameter $\boldsymbol{\theta}$ in which each element is the value of $Y(\boldsymbol{\theta},\mathbf{s}_i)$ at $n$ grid points $\mathbf s_1,\dotsm \mathbf{s}_{n}\in \mathcal{S}$. Our study domain contains $n=176\times112$ spatial locations and covers the entire WAIS.

\subsection{Forward and Reverse Processes and Loss Function}
The diffusion model defines a fixed forward kernel $q(\mathbf{Y}_{t}(\boldsymbol{\theta}) | \mathbf{Y}_{t-1}(\boldsymbol{\theta}))$, which gradually perturbs the data by adding noise, together with a learnable reverse kernel $p_{\boldsymbol{\phi}}(\mathbf{Y}_{t-1}(\boldsymbol{\theta}) | \mathbf{Y}_{t}(\boldsymbol{\theta}))$, which progressively removes the noise and recovers the samples. Specifically, the forward and reverse processes are defined as
\begin{equation}
    \begin{aligned}
        \label{eq: diffusion model process}
        q(\mathbf{Y}_{t}(\boldsymbol{\theta}) | \mathbf{Y}_{t-1}(\boldsymbol{\theta})) &:= N\left(\sqrt{1-\beta_t}\mathbf{Y}_{t-1}(\boldsymbol{\theta}), \beta_t \mathbf{I}\right),\\ 
        p_{\boldsymbol{\phi}}(\mathbf{Y}_{t-1}(\boldsymbol{\theta}) | \mathbf{Y}_{t}(\boldsymbol{\theta})) &:= N\left( \boldsymbol{\mu}_{\boldsymbol{\phi}} (\mathbf{Y}_{t}(\boldsymbol{\theta}),\boldsymbol{\theta}, t), \boldsymbol{\Sigma}_{\boldsymbol{\phi}}(\mathbf{Y}_{t}(\boldsymbol{\theta}),\boldsymbol{\theta},t)\right)
    \end{aligned}
\end{equation}
for $t=1,\dotsm, T$, where $\boldsymbol{\phi}$ is a vector of diffusion model parameters and $\beta_t$ is a noise schedule defined a priori. We define that $\mathbf{Y}_{0}(\boldsymbol{\theta})=\mathbf{Y}(\boldsymbol{\theta})$ and assume that we can approximate $\mathbf{Y}_{T}(\boldsymbol{\theta})=\boldsymbol{\epsilon}\sim N(\mathbf{0}, \mathbf I)$. The mean $\boldsymbol{\mu}_{\boldsymbol{\phi}}(\cdot)$ and variance $\boldsymbol{\Sigma}_{\boldsymbol{\phi}}(\cdot)$ are reconstructed by a neural network. The loss function is 
\begin{equation} 
    \label{eq: loss function 1}
    L_t = \mathbb E_{t,\boldsymbol{\epsilon},\mathbf{Y}(\boldsymbol{\theta})}\left[\|\mathbf{Y}(\boldsymbol{\theta}) - g_{\boldsymbol{\phi}}(\mathbf{Y}_t(\boldsymbol{\theta}), t,\boldsymbol{\theta})\|_2^2 \right],
\end{equation}
where $t\sim\{1,\dotsm, T\}$ and the denoising function $g_{\boldsymbol{\phi}}(\cdot)$ is a neural network such as a U-net \citep{Ronneberger2015}. Following the standard training procedure for the diffusion model \citep{HoEtAl2020}, we repeatedly sample $t$ and $\boldsymbol{\epsilon}$ and update $L_t$ using gradient descent. In \eqref{eq: diffusion model process}, $\boldsymbol{\mu}_{\boldsymbol{\phi}}(\cdot)$ and  $\boldsymbol{\Sigma}_{\boldsymbol{\phi}}(\cdot)$  are reconstructed using $g_{\boldsymbol{\phi}}(\cdot)$ in \eqref{eq: loss function 1}. 

\subsection{Spatial Attention}
A key challenge is that input parameters can affect either the entire spatial domain or specific regions. To account for this spatial heterogeneity, we employ the Adaptive Group Normalization (AdaGN) \citep{dhariwal2021adagn} and decoupled attention \citep{ye2023ipadapter} frameworks for global and local conditioning, respectively, within $g_{\boldsymbol{\phi}}(\cdot)$. For an intermediate feature map $\mathbf{x}\in\mathbb{R}^{C\times H\times W}$, we define the AdaGN layer for global conditioning as: 
\begin{equation}
    \label{eq: adagn}
    \mathrm{AdaGN}(\mathbf{x};\boldsymbol{\theta},t)_c = \{1+\gamma_c(\boldsymbol{\theta},t)\} \mathrm{GroupNorm}(\mathbf{x})_c + \alpha_c(\boldsymbol{\theta},t), \qquad c=1,\dots,C,
\end{equation}
where $\gamma_c(\boldsymbol{\theta},t)$ and $\alpha_c(\boldsymbol{\theta},t)$ are channel-wise scale and shift coefficients generated from an embedding of $(\boldsymbol{\theta},t)$. Because these learning parameters are shared across all spatial locations $H$ and $W$ within a channel, AdaGN provides a global conditioning mechanism. Here, $\mathrm{GroupNorm}(\mathbf{x})_c\in \mathbb R^{H\times W}$ denotes the $c$-th feature map after group normalization, where the normalization is computed over the dimensions $(H, W)$ and across the subset of channels belonging to the same group. 

For local conditioning, we adopt a decoupled attention layer that allows the effects of the input parameters to vary across spatial locations:
\begin{equation}
    \label{eq: decouple attention}
    \mathrm{dAttn}(\mathbf{x}; \boldsymbol{\theta}) = \mathrm{Softmax}\left(\frac{\mathbf{Q}_\mathbf{x}\mathbf{K}_\mathbf{x}^\top}{\sqrt{d}}\right)\mathbf{V}_\mathbf{x} + \mathrm{Softmax}\left(\frac{\mathbf{Q}_\mathbf{x}\mathbf{K}_{\boldsymbol{\theta}}^\top}{\sqrt{d}}\right)\mathbf{V}_{\boldsymbol{\theta}},
\end{equation}
where $(H W,d)$-dimension matrices $\mathbf Q_{\mathbf x}$, $\mathbf K_{\mathbf x}$, and $\mathbf V_{\mathbf x}$ are the query, key, and value projections obtained from the feature map $\mathbf x$, respectively, while $\mathbf K_{\boldsymbol\theta}$ and $\mathbf V_{\boldsymbol\theta}$ are the corresponding key and value projections obtained from the conditioning variable $\boldsymbol\theta$. Here, $d$ denotes the inner dimension of the projected query and key embeddings used to compute the dot-product attention scores. $\mathrm{Softmax}\left(\cdot\right)$ is the row-wise softmax function, which applies the standard softmax normalization to each row of the input matrix. The input parameter $\boldsymbol{\theta}$ is mapped to a collection of ten conditioning tokens, one for each input parameter, from which $\mathbf{K}_{\boldsymbol{\theta}}$ and $\mathbf{V}_{\boldsymbol{\theta}}$ are constructed. The resulting cross-attention weights allow each spatial location in $\mathbf{x}$ to assign different importance to the individual input parameters, thereby providing a location-specific conditioning mechanism.

To integrate global and local conditioning, the U-net architecture $g_{\boldsymbol{\phi}}(\cdot)$ uses a residual block $\mathcal{R}(\cdot)$ and an attention block $\mathcal{A}(\cdot)$. First, the residual block is defined as:
\begin{equation}
    \nonumber
    \begin{aligned}
        \mathbf{F}_1 &= \mathbf{C}_1 \ast \text{SiLU}(\text{GroupNorm}(\mathbf{x})) + \mathbf{B}_1, \\
        \mathcal{R}(\mathbf{x};\boldsymbol{\theta},t) &= \mathbf{x} + \mathbf{C}_2 \ast \text{SiLU}\big(\text{AdaGN}(\mathbf{F}_1; \boldsymbol{\theta}, t)\big) + \mathbf{B}_2,
    \end{aligned}
\end{equation}
where $\ast$ is a convolution operator, $\text{SiLU}(\cdot)$ is a sigmoid linear unit activation function. 
Here, $\mathbf{C}_1\in\mathbb{R}^{C'\times C\times K\times K}$ and $\mathbf{C}_2\in\mathbb{R}^{C'\times C'\times K\times K}$ denote the convolution kernels, while $\mathbf{B}_1,\mathbf{B}_2\in\mathbb{R}^{C'}$ are the corresponding channel-wise bias vectors.
Subsequently, the local conditioning is applied through the attention block using the decoupled attention from \eqref{eq: decouple attention}:
\begin{equation}
    \nonumber
    \mathcal{A}(\mathbf{x}; \boldsymbol{\theta}) = \mathbf{x} + \mathbf{C}_{\text{proj}} \ast \text{dAttn}(\mathbf{x}; \boldsymbol{\theta}) + \mathbf{B}_{\text{proj}}.
\end{equation}
To manage computational complexity, we apply the attention block $\mathcal{A}(\cdot)$ only at selected spatial resolutions and the residual block $\mathcal{R}(\cdot)$ throughout the network. Let $\mathcal{M}_{\text{attn}} \subset \{1, \dots, M\}$ denote the subset of layer indices where local conditioning is active. For the $m$-th layer, the intermediate representation is recursively updated as:
\begin{equation}
    \label{eq: Composition}
    \mathbf{x}^{(m)} = 
    \begin{cases} 
        \mathcal{A}^{(m)}\big(\mathcal{R}^{(m)}(\mathbf{x}^{(m-1)}; \boldsymbol{\theta}, t); \boldsymbol{\theta}\big), & \text{if } m \in \mathcal{M}_{\text{attn}}, \\ 
        \mathcal{R}^{(m)}(\mathbf{x}^{(m-1)}; \boldsymbol{\theta}, t), & \text{otherwise},
    \end{cases}
\end{equation}
for $m=1,\dotsm, M$. By repeating this sequential structure through the U-net $g_{\boldsymbol{\phi}}(\cdot)$, the network hierarchically injects the information of $\boldsymbol{\theta}$ at every resolution, capturing domain-wide global effects first by $\mathcal{R}(\cdot)$ and iteratively refining spatially varying local interactions using $\mathcal{A}(\cdot)$. 

\subsection{Bit Diffusion for Handling Zeros}

The excess zeros in the ice patterns prevent the direct application of \eqref{eq: diffusion model process}, because the reverse process relies on a Gaussian approximation. To address this, we introduce a latent variable $\Gamma_Y(\boldsymbol{\theta},\mathbf{s}_i)$, which represents the ice presence or absence pattern, such that
\begin{equation}
    \nonumber
    \Gamma_Y(\boldsymbol{\theta},\mathbf{s}_i)=\begin{cases}
        1, & \text{if } Y(\boldsymbol{\theta},\mathbf{s}_i)>0,\\
        -1, & \text{if } Y(\boldsymbol{\theta},\mathbf{s}_i) = 0,
\end{cases}
\quad i=1,\dotsm, n.
\end{equation}
Here, the sign of $\Gamma_Y(\boldsymbol{\theta},\mathbf{s}_i)$ indicates the presence or absence of the ice. \cite{Chen2022AnalogBG} introduced a framework for modeling discrete data through Gaussian diffusion by leveraging binary bit representations and thresholding. Motivated by their construction, we model the latent variable $\Gamma_Y(\boldsymbol{\theta},\mathbf{s}_i)$ through the same diffusion architecture for continuous outcomes in \eqref{eq: diffusion model process}, and it can be converted to a discrete state by thresholding at zero. 

Let $\mathbf{Y}^*(\boldsymbol{\theta}) = \left[ \mathbf{Y}^*(\boldsymbol{\theta},\mathbf{s}_1) ,\dots,\mathbf{Y}^*(\boldsymbol{\theta},\mathbf{s}_n) \right]$ be a collection of the bivariate vector $\mathbf{Y}^*(\boldsymbol{\theta},\mathbf{s}_i) = [Y(\boldsymbol{\theta},\mathbf{s}_i),\Gamma_Y(\boldsymbol{\theta},\mathbf{s}_i)]$ defined at each grid point $\mathbf{s}_i\in \mathcal{S}$. Rather than constructing separate emulators for individual elements  $Y(\boldsymbol{\theta},\mathbf{s}_i)$ and $\Gamma_Y(\boldsymbol{\theta},\mathbf{s}_i)$, we use a single conditional diffusion model to jointly generate both variables for a given $\boldsymbol{\theta}$. By replacing the $\mathbf{Y}(\boldsymbol{\theta})$ in \eqref{eq: loss function 1} with $\mathbf{Y}^*(\boldsymbol{\theta})$, we train the generative model  $g_{\boldsymbol{\phi}}$ using the loss function 
\begin{equation}
    \nonumber
    L_D^* = \mathbb E_{t,\boldsymbol{\epsilon},\mathbf{Y}^*(\boldsymbol{\theta})}\left[\|\mathbf{Y}^*(\boldsymbol{\theta}) - g_{\boldsymbol{\phi}}(\mathbf{Y}^*_t(\boldsymbol{\theta}), t,\boldsymbol{\theta})\|_2^2 \right]. 
\end{equation}
Once $g_{\boldsymbol{\phi}}$ is trained, predicted outputs can be generated through the Markovian reverse process in \eqref{eq: diffusion model process}. However, running all $T$ steps is computationally prohibitive in practice. To accelerate this process, we employ the denoising diffusion implicit model (DDIM) \citep{SongDDIM2022}, which relaxes the Markovian assumption and allows for deterministic sampling over a much shorter sub-sequence of timesteps $\tau_S> \tau_{S-1}> \dots> \tau_1$ where $S \ll T$. Starting from a bivariate initial random noise $\boldsymbol{\epsilon}^*$, the joint state of continuous and discrete latent variables is recursively updated using the predictions of $g_{\widehat{\boldsymbol{\phi}}}$ with a set of trained model parameters $\boldsymbol{\widehat{{\phi}}}$. Through this iterative sampling process, we can generate predicted values for $\mathbf{Y}^*(\boldsymbol{\theta})$. 

The emulator must output a univariate ice-thickness field with exact zeros in ice-free regions. Let $\widehat{Y}(\boldsymbol{\theta}, \mathbf{s}_i)$ and $\widehat{\Gamma}_Y(\boldsymbol{\theta}, \mathbf{s}_i)$ be the predicted ice thickness and latent variables at location $\mathbf{s}_i$, respectively. 
We define the emulator output $\boldsymbol{\eta}(\boldsymbol{\theta})$ in which the value at the $i$-th location is computed by
\begin{equation}
    \nonumber
    \eta(\boldsymbol{\theta}, \mathbf{s}_i) = \widehat{Y}(\boldsymbol{\theta},\mathbf{s}_i)\mathbb{I}\{\widehat{\Gamma}_Y(\boldsymbol{\theta}, \mathbf{s}_i)\geq 0\},\quad i=1,\dotsm,n,
\end{equation}
where $\mathbb{I}(\cdot)$ is an indicator function. Since we assume that the sign of $\Gamma_Y(\boldsymbol{\theta},\mathbf{s}_i)$ encodes the binary ice pattern, the field $\mathbb I\{\widehat{\Gamma}_Y(\boldsymbol{\theta},\mathbf{s}_i)\geq 0\}$ determines ice presence. Multiplying this indicator by the generated thickness field enforces exact zeros in ice-free regions while retaining the generated thickness values where ice is present. This representation transforms predictions for $\mathbf{Y}^*(\boldsymbol{\theta})$ into the ice pattern $\boldsymbol{\eta}(\boldsymbol{\theta})$. With multiple draws of $\boldsymbol{\epsilon}^*$, we can generate Monte Carlo samples of emulated outputs.

\section{Likelihood-free Calibration}
\label{sec: Likelihood-Free Calibration}
Following the emulation step, calibration is performed to infer $\boldsymbol{\theta}$ from observational data. Let $\mathbf{Z}=(Z(\mathbf{s}_1), \dotsc, Z(\mathbf{s}_n))^\top$ be the $n$-dimensional vectorized observational data observed at the same locations $\mathbf{s}_1,\dotsc, \mathbf{s}_n \in\mathcal{S}$ as $\mathbf{Y}(\boldsymbol{\theta})$ and $\boldsymbol{\eta}(\boldsymbol{\theta})$. Note that standard likelihood-based calibration approaches \citep{KennedyandHagun2002} are not directly applicable because our emulator provides samples from an intractable distribution \citep{Li2023}. A conventional approach to circumventing this issue is the ABC rejection method \citep{Marinabc}, which draws a proposal $\boldsymbol{\theta}'$ from the prior distribution $p(\boldsymbol{\theta})$, generates the corresponding candidate $\boldsymbol{\eta}(\boldsymbol{\theta}')$, and accepts $\boldsymbol{\theta}'$ if $d(\boldsymbol{\eta}(\boldsymbol{\theta}'),\mathbf Z)<\varepsilon$, where $d(\cdot,\cdot)$ is a distance measure and $\varepsilon>0$ is a tolerance. The principle behind this framework is to approximate the likelihood $p(\mathbf Z|\boldsymbol{\theta})$ using $p(d(\boldsymbol{\eta}(\boldsymbol{\theta}),\mathbf Z)<\varepsilon|\boldsymbol{\theta})$ for sufficiently small $\varepsilon$.

However, the performance of ABC depends on the choice of $d(\cdot,\cdot)$ and $\varepsilon$. A small value of $\varepsilon$ can yield prohibitively low acceptance rates, especially when $p(\boldsymbol{\theta})$ is far from the posterior distribution, while a large $\varepsilon$ or an uninformative construction of $d(\cdot,\cdot)$ can lead to poor posterior approximation. These difficulties are further amplified for high-dimensional data, where defining an appropriate $d(\cdot,\cdot)$ is challenging. Although low-dimensional summary statistics are often used to mitigate this issue \citep{Marinabc}, identifying informative and low-dimensional summary statistics remains nontrivial in complex computer model calibration problems.

In this section, we introduce the calibration component of SC-DS, a likelihood-free approach based on a Siamese network. The Siamese network is trained to compute the probability that two ice patterns are generated from the same parameter setting of the PSU3D-ICE model. In this way, the Siamese network can compare two ice patterns based on features highly relevant to parameter calibration. By using this metric as a probabilistic acceptance criterion, our method avoids the need to specify a distance measure and a tolerance, which are substantial challenges in traditional ABC-based approaches. To further improve computational efficiency, we employ a sequential sampling approach that utilizes a series of intermediate distributions, with the Siamese network retrained at each iteration.

In addition to this computational consideration, we also propose a procedure to handle data–model discrepancy. By downweighting the spatial locations with low model accuracy during Siamese network training, we encourage the calibration algorithm to focus on the spatial features that are accurately captured by the PSU3D-ICE model.

\subsection{Likelihood-free Calibration Approach}
\label{sec: Likelihood-free calibration approach}
We first describe the overall calibration procedure. The main idea is to replace the deterministic thresholding rule, which accepts the proposed parameter $\boldsymbol{\theta}'\sim p(\boldsymbol{\theta})$ if $d(\boldsymbol{\eta}(\boldsymbol{\theta}'),\mathbf Z)<\varepsilon$ with a probabilistic acceptance rule defined by a deep neural network. Let $\rho(\boldsymbol{\eta}(\boldsymbol{\theta}),\mathbf{Z})\in[0,1]$ denote the output of the Siamese network that compares the emulated output $\boldsymbol{\eta}(\boldsymbol{\theta})$ with the observational data $\mathbf{Z}$. We interpret the output $\rho(\cdot,\cdot)$ as a similarity rather than a distance, so larger values indicate stronger agreement between the two inputs. Given $\rho(\cdot,\cdot)$, we replace the deterministic rule with the following acceptance criterion:
\begin{equation}
    \label{eq: acceptance rule}
    \text{accept proposed } \boldsymbol{\theta}'\sim p(\boldsymbol{\theta}) \text{ with probability }\rho(\boldsymbol{\eta}(\boldsymbol{\theta}'),\mathbf{Z}).
\end{equation}
Using this criterion, we can replace the manually specified pair $(d(\cdot,\cdot),\varepsilon)$ in standard ABC with the neural-network-based acceptance probability $\rho(\cdot,\cdot)$. The detailed construction of $\rho(\cdot,\cdot)$ is described in Section~\ref{sec: Architecture of the Siamese network}.
 
However, applying the probabilistic rule in \eqref{eq: acceptance rule} can still be inefficient when the prior distribution $p(\boldsymbol{\theta})$ places little mass near parameter values that generate outputs similar to the observation. This situation is common in calibration problems, since the prior is typically a multivariate uniform distribution spanning the physically plausible ranges of the input parameters. To address this issue, various sequential Monte Carlo approaches, which propagate the sample distribution toward regions of high posterior mass, have been developed for model calibration problems \citep{ToniABCSMC, JeremiahSissonSMC, Lee2020}. Inspired by them, we incorporate a method that gradually guides parameter proposals toward regions with large values of $\rho(\boldsymbol{\eta}(\boldsymbol{\theta}),\mathbf{Z})$. In addition, we introduce a retraining strategy that updates the Siamese neural network for $\rho(\cdot,\cdot)$ using samples generated from the sequential populations. This adaptation is useful because distinguishing between samples becomes more difficult as the proposal distribution concentrates around observation-like outputs. By retraining $\rho(\cdot,\cdot)$ on progressively refined samples, the acceptance criterion can better distinguish subtle differences among highly similar samples and becomes sharper in later stages.
 
\begin{enumerate}
    \item \emph{Initialization}: Train $\rho(\cdot,\cdot)$ using emulator outputs generated at an initial set of parameter samples.
    Set $s=0$ and $i=1$.
    \item \emph{Sampling from the proposal distribution}: If $s=0$, sample $\boldsymbol{\theta}''$ from the prior $p(\boldsymbol{\theta})$.\\
    Else, sample $\boldsymbol{\theta}'$ from the previous samples $\{\boldsymbol{\theta}_i^{(s-1)}\}_{i=1}^N$ with probabilities proportional to $\{\rho(\boldsymbol{\eta}(\boldsymbol{\theta}_i^{(s-1)}),\mathbf{Z})\}_{i=1}^N$ and perturb it by adding Gaussian noise to get $\boldsymbol{\theta}''$ (i.e., $\boldsymbol{\theta}'' \sim N(\boldsymbol{\theta}',\sigma^2 \mathbf I)$).
    \item \emph{Generate an emulated output}: If $p(\boldsymbol{\theta}'')=0$, return to Step 2.\\
    Else, generate $\boldsymbol{\eta}(\boldsymbol{\theta}'')$ from the emulator.
    \item \emph{Compute the similarity}: Compute the similarity $\rho(\boldsymbol{\eta}(\boldsymbol{\theta}''),\mathbf{Z})$ using the Siamese network.
    \item \emph{Accept}: With probability $\rho(\boldsymbol{\eta}(\boldsymbol{\theta}''),\mathbf{Z})$, accept the proposal by setting $\boldsymbol{\theta}_i^{(s)} = \boldsymbol{\theta}''$ and $i \leftarrow i + 1$; otherwise reject it. In either case, if $i \le N$, return to Step 2.
    \item \emph{Check convergence}: If $s \ge 1$ and the distribution of the accepted particles $\{\boldsymbol{\theta}_i^{(s)}\}_{i=1}^N$ is sufficiently close to that of the previous stage $\{\boldsymbol{\theta}_i^{(s-1)}\}_{i=1}^N$, terminate the procedure.
   \item \emph{Retrain the Siamese network}: Otherwise, retrain the Siamese network for  $\rho(\cdot,\cdot)$ using emulated samples from $\{\boldsymbol{\theta}_i^{(s)}\}_{i=1}^N$, recompute $\{\rho(\boldsymbol{\eta}(\boldsymbol{\theta}_i^{(s)}),\mathbf{Z})\}_{i=1}^N$ using the retrained $\rho(\cdot,\cdot)$, set $s \leftarrow s+1$ and $i \leftarrow 1$, and return to Step 2.
\end{enumerate}

The dataset for the initial training of $\rho(\cdot,\cdot)$ in Step 1 is constructed by generating an initial set of parameter samples via a Latin hypercube design within the [0,1] interval and computing the corresponding emulator outputs. The perturbation variance $\sigma^2$ is set to $0.1^2$ to limit the probability of proposing values outside the normalized support $[0,1]^{10}$. We set $N=1000$ at each stage to allow the intermediate sample distributions to sufficiently explore the parameter space and to prevent particle degeneracy caused by a small number of samples with dominant weights. To exploit GPU-based parallel computation, Steps 2--5 are executed in batches: at each iteration, a batch of $B$ candidate parameters is proposed, emulated, and scored simultaneously, and the acceptance decisions of Step 5 are applied independently within the batch. Accepted particles are accumulated across successive batches until at least $N$ accepted parameter values have been collected, at which point the population $\{\boldsymbol{\theta}_i^{(s)}\}_{i=1}^N$ is formed from the first $N$ accepted particles and the stage proceeds to the convergence check (Step 6). Convergence is assessed by comparing the distributions of the previous and current particles ($\{\boldsymbol{\theta}_i^{(s-1)}\}_{i=1}^N$ and $\{\boldsymbol{\theta}_i^{(s)}\}_{i=1}^N$). If the two distributions are sufficiently close, the procedure terminates; otherwise, the Siamese network is retrained (Step 7), and the resampling steps are repeated.

\subsection{Architecture of the Siamese Network}
\label{sec: Architecture of the Siamese network}
This subsection describes the Siamese network used to compute the acceptance probability $\rho(\cdot,\cdot)$. The network consists of two identical, weight-sharing convolutional neural networks (CNNs) that map input fields, denoted generically by $\mathbf{U}$ and $\mathbf{V}$, into low-dimensional feature vectors. These vectors are then combined to yield a scalar similarity score. We first describe the overall architecture of the Siamese network, and then the specific choices of $(\mathbf{U}, \mathbf{V})$ during the training and calibration stages.
 
\begin{figure}[htbp]
    \centering
    \IfFileExists{plot/SiameseArchitecture.png}{%
        \includegraphics[width=\linewidth]{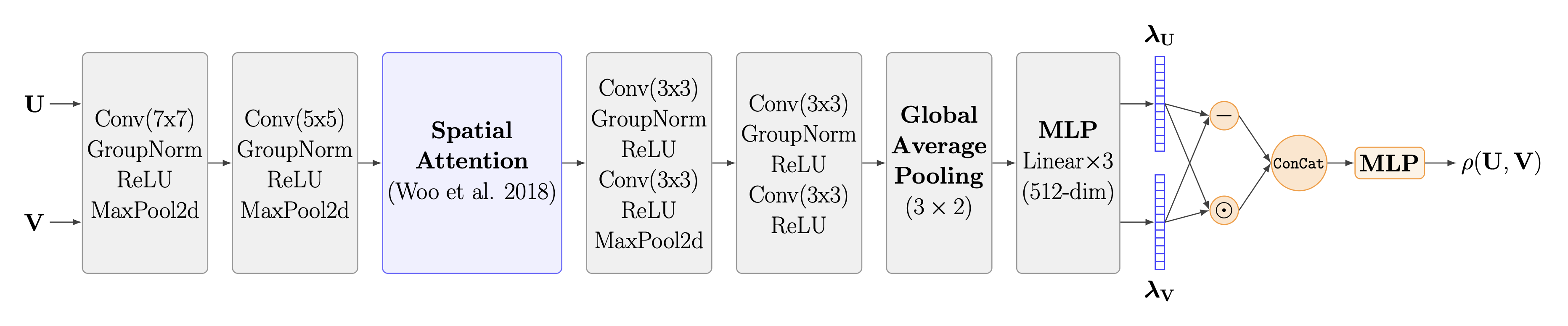}}{%
        \fbox{\parbox{0.9\linewidth}{\centering\vspace{2.5cm}%
        [figure placeholder: plot/SiameseArchitecture.png]\vspace{2.5cm}}}}
    \caption{Architecture of the Siamese network used to compute $\rho(\mathbf{U},\mathbf{V})$. Two identical CNNs map input fields to a low-dimensional feature space, which is then combined through symmetric operations and a final multilayer perceptron.}
    \label{fig: Siamese architecture}
\end{figure}

Figure~\ref{fig: Siamese architecture} summarizes the architecture of the Siamese network used in our study. Each input field, either $\mathbf{U}$ or $\mathbf{V}$, is first passed through a sequence of convolutional blocks, where each block consists of a convolution layer, group normalization, a ReLU activation, and, when specified, a max-pooling operation to progressively reduce the spatial resolution while increasing the receptive field. A spatial attention layer \citep{woo2018cbam} is inserted to weight spatial locations based on their importance in distinguishing ice patterns generated under different parameter settings. The resulting feature maps are processed by additional convolutional blocks, aggregated into a single feature vector by global average pooling, and finally transformed by fully connected layers into a $q$-dimensional representation. Detailed architectural specifications are provided in Section~C of the Supplementary Material.
 
Let $\boldsymbol{\lambda}_{\mathbf{U}}$ and $\boldsymbol{\lambda}_{\mathbf{V}}$ be the $q$-dimensional feature vectors produced by the two CNN components of the Siamese network when the inputs are $\mathbf{U}$ and $\mathbf{V}$, respectively. To obtain a similarity measure between $\mathbf{U}$ and $\mathbf{V}$, we compute the $2q$-dimensional vector, 
    $\boldsymbol{\omega} = \left(|\boldsymbol{\lambda}_{\mathbf{U}} -\boldsymbol{\lambda}_{\mathbf{V}}|,
    \boldsymbol{\lambda}_{\mathbf{U}} \odot \boldsymbol{\lambda}_{\mathbf{V}} \right)^\top$,
which stacks the element-wise absolute difference and the element-wise product of the two feature vectors. Both operations are symmetric in $\mathbf{U}$ and $\mathbf{V}$, so the resulting similarity is invariant to the order of the inputs. We compute the similarity score $\rho(\cdot,\cdot)$ as
\begin{equation}
    \label{eq: acceptance probability}
    \rho(\mathbf{U},\mathbf{V}) = \sigma(\boldsymbol{\beta} \boldsymbol{\omega} + \alpha)
    =\frac{1}{1 + \exp\{-(\boldsymbol{\beta} \boldsymbol{\omega} +\alpha)\}},
\end{equation}
where $\boldsymbol{\beta}\in\mathbb R^{1\times 2q}$ and $\alpha\in\mathbb R$ are the weight vector and bias of the linear layer that projects $\boldsymbol{\omega}$ to a scalar. The sigmoid function $\sigma(\cdot)$ in \eqref{eq: acceptance probability} then maps this scalar to the interval $(0,1)$.
 
The same network uses different input pairs at the training and calibration stages. During training, it acts as a binary classifier for matched and unmatched pairs. To be more specific, $(\mathbf{U},\mathbf{V})$ is a pair of emulator outputs $(\boldsymbol{\eta}(\boldsymbol{\theta}_j),\boldsymbol{\eta}(\boldsymbol{\theta}_{j'}))$, and the network is trained to assign a higher $\rho$ to matched pairs generated from the same parameter setting ($\boldsymbol{\theta}_j=\boldsymbol{\theta}_{j'}$) and a lower $\rho$ to unmatched pairs from different settings ($\boldsymbol{\theta}_j\ne\boldsymbol{\theta}_{j'}$). In the calibration stage, the output of the network is used to compute the probability of accepting a proposed parameter value. That is, $(\mathbf{U},\mathbf{V})=(\boldsymbol{\eta}(\boldsymbol{\theta}''),\mathbf{Z})$, where $\boldsymbol{\eta}(\boldsymbol{\theta}'')$ is the emulated output for a proposed parameter and $\mathbf{Z}$ is the observed field (Step~4 of Section~\ref{sec: Likelihood-free calibration approach}); the network then returns a higher $\rho$ when the emulated output resembles the observation.
 
Training requires a dataset containing both (i) pairs of emulator outputs generated with the same input parameter setting (matched pairs) and (ii) pairs of emulator outputs from different parameter settings (unmatched pairs). To this end, we sample $\boldsymbol{\theta}_1,\dots, \boldsymbol{\theta}_J$ using a Latin hypercube design and generate $K$ emulator outputs $\boldsymbol{\eta}_1(\boldsymbol{\theta}_j),\dots, \boldsymbol{\eta}_K(\boldsymbol{\theta}_j)$ for each $\boldsymbol{\theta}_{j}$. This yields $J\binom{K}{2}$ pairs of emulator outputs corresponding to matched input parameters and $K^2\binom{J}{2}$ pairs corresponding to unmatched input parameters. For balanced training, we construct the training and validation datasets by sampling an equal number of pairs from each class, up to the smaller of $J\binom{K}{2}$ and $K^2\binom{J}{2}$. Let $\mathcal{D}$ be the set of sampled index tuples $(j,k,j',k')$ that identify the selected pairs of emulator outputs. The network is trained by minimizing the binary cross-entropy loss 
\begin{equation}
    \label{eq: siamese loss function}
    L_S = -\frac{1}{|\mathcal{D}|} \sum_{(j,k,j',k')\in \mathcal{D}} \left[\mathbb I(j=j')\log \rho(\boldsymbol{\eta}_k(\boldsymbol{\theta}_j), \boldsymbol{\eta}_{k'}(\boldsymbol{\theta}_{j'})) + \mathbb I(j \ne j') \log\{1- \rho(\boldsymbol{\eta}_k(\boldsymbol{\theta}_j), \boldsymbol{\eta}_{k'}(\boldsymbol{\theta}_{j'}))\}\right],
\end{equation}
where $|\mathcal{D}|$ denotes the number of elements in $\mathcal{D}$.

\subsection{Data-model Discrepancy}
\label{sec: Data-model discrepancy}
Even when the emulator and the Siamese network are available, the proposed calibration step may not work well due to inadequacies in the computer model. Typically, owing to data-model discrepancy \citep{Brynjarsdottir_2014}, certain small areas of the emulator output $\boldsymbol{\eta}(\boldsymbol{\theta})$ can persistently differ from the observation $\mathbf{Z}$ even for the best values of $\boldsymbol{\theta}$, i.e., the values leading to a very good overall match between $\boldsymbol{\eta}(\boldsymbol{\theta})$ and $\mathbf{Z}$ everywhere except in those areas. This can cause the proposed procedure to reject almost all proposed $\boldsymbol{\theta}''$ values because $\rho(\boldsymbol{\eta}(\boldsymbol{\theta}''),\mathbf{Z})$ remains small due to those mismatches.

To incorporate potential discrepancies, we shrink the model output and the observed data toward zero at the locations where they exhibit notable discrepancies. To be more specific, we train the Siamese network using $\boldsymbol{\delta} \odot \boldsymbol{\eta}(\boldsymbol{\theta})$ where $\boldsymbol{\delta}=(\delta(\mathbf{s}_1),\dotsm, \delta(\mathbf{s}_n))^\top$ takes values close to one in regions where the data-model discrepancy is small and values close to zero in regions where the discrepancy is large. The acceptance probability for the calibration is then computed as $\rho(\boldsymbol{\delta}\odot\boldsymbol{\eta}(\boldsymbol{\theta}''),\boldsymbol{\delta}\odot\mathbf{Z})$. Through this, the Siamese network places greater emphasis on regions where the model can accurately reproduce the observed ice pattern. 

To define $\boldsymbol{\delta}$, we first select the $Q$ best design points and their corresponding model runs according to the mean absolute error (MAE) relative to the observational data, so that the identified discrepancy patterns are derived from model runs that agree well with the observations. Let $\boldsymbol{\theta}^*_q$ denote the $q$-th best design point, and let $Y(\boldsymbol{\theta}^*_q,\mathbf{s}_i)$ denote the value of the corresponding model output at location $\mathbf{s}_i$, for $q=1,\dotsm,Q$ and $i=1,\dotsm, n$. Using these selected model runs, we compute the pointwise mean $\overline{Y}^*(\mathbf{s}_i)$ such that 
\begin{equation}
    \nonumber
    \overline{Y}^*(\mathbf{s}_i) = \frac{1}{Q}\sum_{q=1}^Q Y(\boldsymbol{\theta}^*_q,\mathbf{s}_i),\quad i=1,\dotsm, n
\end{equation}
and define the weight ${\delta}(\mathbf{s}_i)$ as follows: 
\begin{equation}
    \label{eq: discrepancy definition}
    \delta(\mathbf{s}_i) = \max\left[ 0 , 2\min\left\{ \frac{Z(\mathbf{s}_i)}{\overline{Y}^*(\mathbf{s}_i)+c} ,  \frac{\overline{Y}^*(\mathbf{s}_i)}{ Z(\mathbf{s}_i)+c} \right\} - 1 \right],\quad i=1,\dotsm, n,
\end{equation}
where $c>0$ is a small constant introduced to avoid division by zero. This definition gives $\delta(\mathbf{s}_i)\approx1$ when 
$Z(\mathbf{s}_i)=\overline{Y}^*(\mathbf{s}_i)$, and assigns smaller values as the discrepancy between the observations and the pointwise mean of the best model runs increases. In particular, $\delta(\mathbf{s}_i)=0$ when one value is at least twice as large as the other.

In the application problem in Section \ref{sec: Results for Observational Data}, we set $Q=49$, which corresponds to approximately $10\%$ of the $499$ design points. Our empirical results show that this specification of $\boldsymbol{\delta}$ identifies scientifically interpretable data-model discrepancy patterns and directs the calibration procedure to disregard the features that cannot be captured by the PSU3D-ICE model, even at the best input parameter settings (see Section \ref{sec: Results for Observational Data} for further details).

\section{Applications}
\label{sec: Applications}
We now turn to the application of SC-DS to the WAIS model calibration problem described in Section \ref{sec:computer-model-and-data-description}. In Section~\ref{sec: Comparing with Existing Method}, we compare SC-DS with the state-of-the-art method \citep{Chang2022}. For this comparison, we consider a small subset of the model output on the $37 \times 86$ grid used in \cite{Chang2022} because of the computational limitations of the competing method. We then show that SC-DS can use the full data on the $176 \times 112$ grid covering the entire WAIS, as shown in Figure~\ref{fig: Examples and observational data}. Specifically, in Section~\ref{sec: Diagnostics of the Performance in Large Scale Model Outputs}, we first diagnose the emulator's predictive performance on held-out model runs and then present calibration results for one representative held-out run. Section~\ref{sec: Results for Observational Data} presents the calibration results for the real observational data.
 
Note that the discrepancy representation in Section~\ref{sec: Data-model discrepancy} is applied only to the observational data analysis in Section~\ref{sec: Results for Observational Data}. We do not apply it in the other two experiments, where the data--model discrepancy is either not large enough to affect performance (Section~\ref{sec: Comparing with Existing Method}) or is absent because one of the original model runs is used as the target observation (Section~\ref{sec: Diagnostics of the Performance in Large Scale Model Outputs}). We perform calibration for all ten input parameters, but we report results for four parameters in this article: the ocean sub-ice-shelf melt factor (OCFAC), the calving factor (CALV), the basal sliding coefficient (CRH), and the asthenospheric relaxation e-folding time (TAU), which are glaciologically more important than the other six input parameters \citep{Chang2022}. The results for the remaining input parameters are shown in Section~E of the Supplementary Material. SC-DS was implemented in Python and executed on an NVIDIA RTX A6000 GPU, and the competing method was implemented in R and executed on a single core of an Intel Xeon Gold processor.

\subsection{Comparison with Existing Method}
\label{sec: Comparing with Existing Method}
In this section, we compare SC-DS with the GP and MCMC-based approach of \citet{Chang2022}, hereafter the GP-based baseline. It is the state-of-the-art ice model calibration method that uses spatial patterns of ice thickness. This method uses emulation and calibration models that, through a likelihood function based on semi-continuous spatial distributions, carry out inference via Markov chain Monte Carlo (MCMC). While the method enables accurate inference for the input parameters of an ice model, its computational cost grows quadratically with the size of the model output and observational data; hence, it does not scale well. Even with a reduced model output on a $37 \times 86$ grid, the method requires tens of hours to ensure proper mixing of the MCMC chain. This precludes its application to the full model output on the $176 \times 112$ grid.
 
We compare the two calibration methods on the synthetic observational data used in \cite{Chang2022}, for which the GP-based baseline is computationally affordable. Here, a model run resembling the real observed ice pattern is selected as the ground truth, and artificial errors are added to this output to mimic plausible discrepancies between the model and reality; see \cite{Chang2022} for further details. The data region spans $37 \times 86$ pixels located in the Amundsen Sea Embayment (ASE), which is expected to be one of the regions most susceptible to climate change \citep{Chang2016,Chang2022}.
 
\begin{figure}[htbp]
\centering
\IfFileExists{plot/gp_compare/contour_gp_simple.png}{%
  \includegraphics[width=0.48\linewidth]{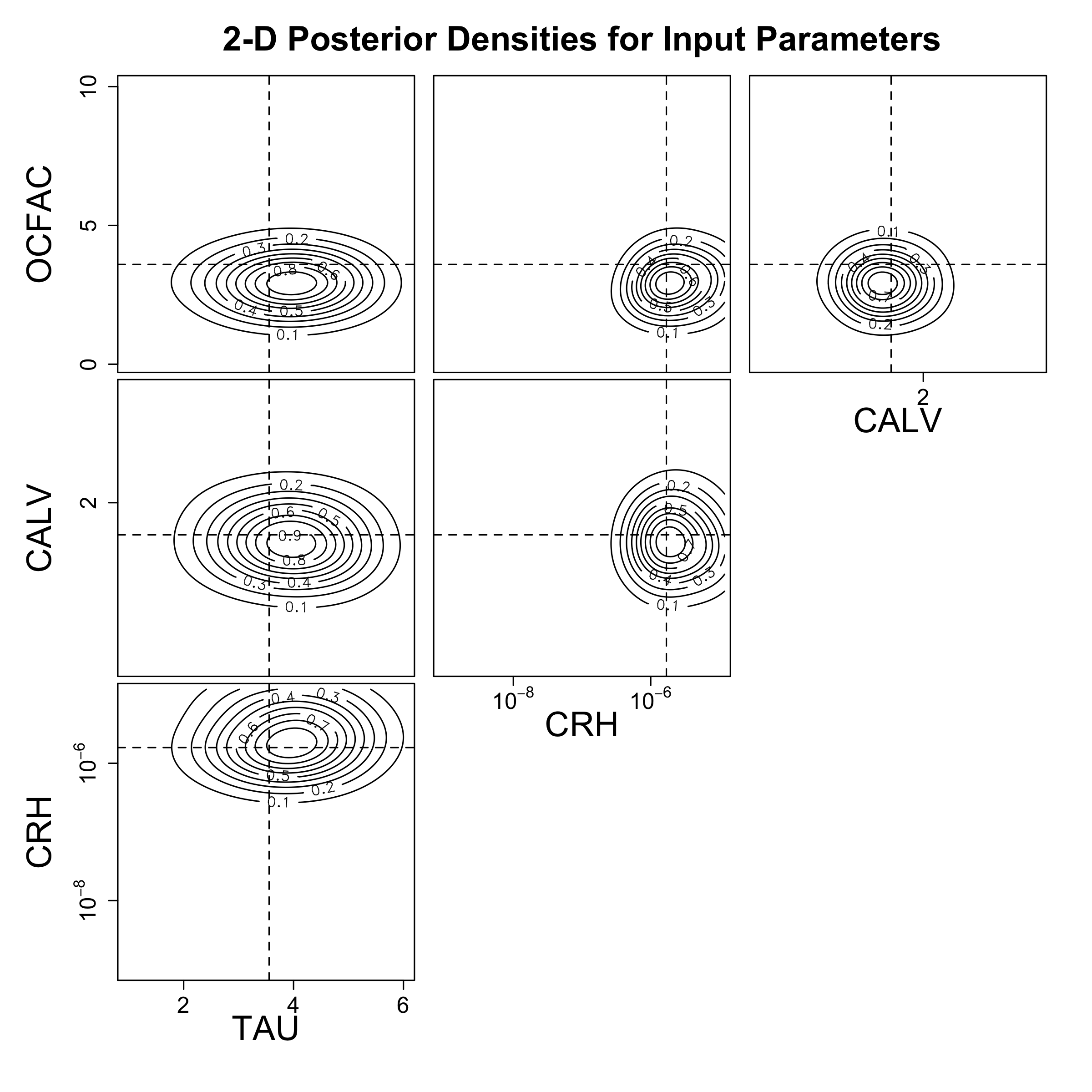}%
  \includegraphics[width=0.48\linewidth]{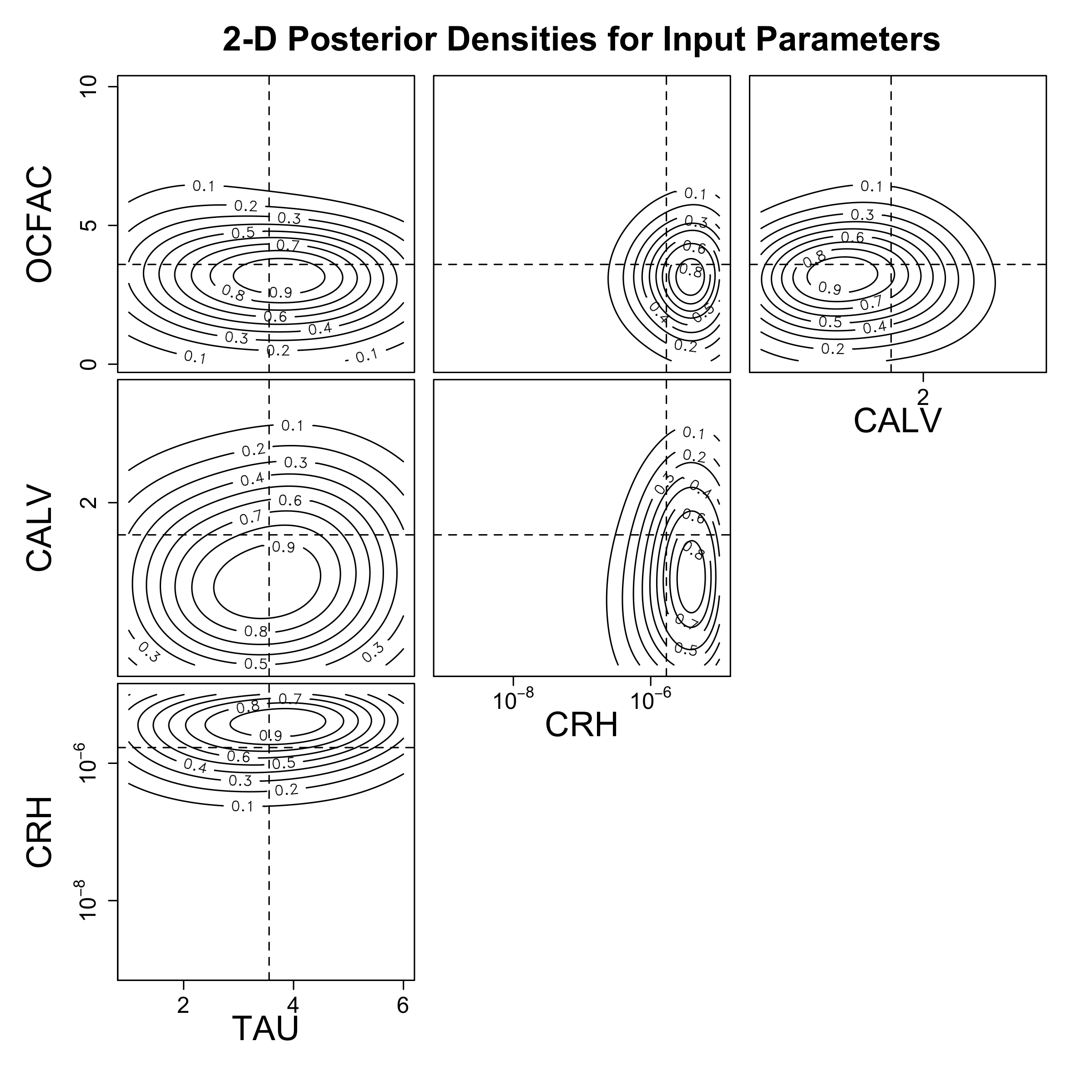}}{%
  \fbox{\parbox{0.9\linewidth}{\centering\vspace{2cm}[figure placeholder: GP contour (left) / our contour (right)]\vspace{2cm}}}}
\caption{The contour plots of the GP-based baseline (left) and SC-DS (right). The dashed lines show the input parameter settings for the synthetic truth. The plots for every input parameter are visualized in Section~E of the Supplementary Material.}
\label{fig: ContourComparing}
\end{figure}

For the GP-based baseline, we obtain an MCMC chain of 2,000,000 iterations, which converges to equilibrium within the first half and remains stable thereafter. The overall computing time is 198.9 hours with an R implementation. For SC-DS, the sequential procedure converges after a total of four stages, including the initial stage in which the particles are sampled from the prior distribution. Each population consists of $N=1,000$ particles and is computed in parallel on a GPU with a batch size of 100. We use a uniform prior over each of the ten input parameters, scaled to $[0,1]$. The standard deviation of the perturbation ($\sigma$ in Section \ref{sec: Likelihood-Free Calibration}) is set to $0.1$ to reduce the chance of proposing values outside the support. The calibration requires 2.22 hours of computation. The resulting contour plots for the two methods are visualized in Figure~\ref{fig: ContourComparing}.
 
The bivariate marginal density plots suggest that both methods yield posterior densities centered on the true input parameter setting, while our method exhibits slightly higher uncertainty than the GP-based baseline. This indicates that our likelihood-free approach incurs some information loss compared to the likelihood-based approach. However, the computational gain is substantial even in this small-scale example.
 
We validate the calibration result from the proposed approach by assessing whether the model runs corresponding to posterior samples resemble the target observations. Since running the actual PSU3D-ICE model for the 1,000 particles is computationally expensive, we instead generate 1,000 predicted model runs, one for each particle, from our emulator and compare their pointwise mean with the target synthetic observational data. The MAE of the mean thickness pattern, computed over the locations where the target ice thickness is positive, is approximately 58.49~m for SC-DS and 188.57~m for the GP-based baseline; for reference, the mean of the target thickness pattern over those locations is 2,085.97~m. 

Notably, although SC-DS yields somewhat wider marginal posteriors than the GP-based baseline, it produces a substantially lower MAE for the predicted ice pattern. One possible explanation for this contrast is that predictive accuracy depends not only on posterior uncertainty but also on whether influential parameters are concentrated near their true values. In particular, CRHASE, the multiplicative factor that modifies the basal sliding coefficient in the inshore ASE, is especially relevant to the study domain. As shown in Section~E of the Supplementary Material, the GP-based posterior for CRHASE is centered away from the synthetic truth, whereas the SC-DS posterior is concentrated near it. Because the study domain lies within the ASE, this difference may explain the lower MAE of SC-DS despite its wider posteriors for several other parameters.

To assess the prediction performance for the ice presence--absence pattern, we compute the pointwise mean of the 1,000 predicted binary patterns and then apply a threshold of 0.5, assigning one (i.e., ice presence) if the mean exceeds 0.5 and zero (i.e., ice absence) otherwise. Our method achieves a binary accuracy of 98.4\%, comparable to the GP-based baseline's 98.7\%. Table~\ref{tab: Comparing calibration performance} summarizes the comparison. Maps of the target synthetic data and the emulated outputs from the proposed method and the GP-based baseline are shown in Section~D of the Supplementary Material.

\begin{table}[htbp]
\centering
\resizebox{\textwidth}{!}{%
\begin{tabular}{lccccc}
\toprule
\multicolumn{2}{c}{\textbf{Model Info}} & \multicolumn{2}{c}{\textbf{Calibration Step}} & \multicolumn{2}{c}{\textbf{Validation}} \\
\cmidrule(lr){1-2} \cmidrule(lr){3-4} \cmidrule(lr){5-6}
\textbf{Methods} & \textbf{Device} & \textbf{Computation Time} (hours) & \textbf{Iterations} & \textbf{MAE} (m) & \textbf{Binary Accuracy (\%)} \\
\midrule
GP-based baseline & CPU & 198.9  & 2M iterations  &  188.57  & 98.7 \\
SC-DS & GPU & 2.22  & 4 populations, each 1,000 particles &  58.49  & 98.4 \\
\bottomrule
\end{tabular}
}
\caption{Comparison of calibration and validation performance between the GP-based baseline and SC-DS for the synthetic data.}
\label{tab: Comparing calibration performance}
\end{table}

\subsection{Diagnostics of the Performance in Large Scale Model Outputs}
\label{sec: Diagnostics of the Performance in Large Scale Model Outputs}
\begin{figure}[htbp]
    \centering
    \IfFileExists{plot/emulation_performance/Target_Mean2.png}{%
      \includegraphics[width=0.48\linewidth]{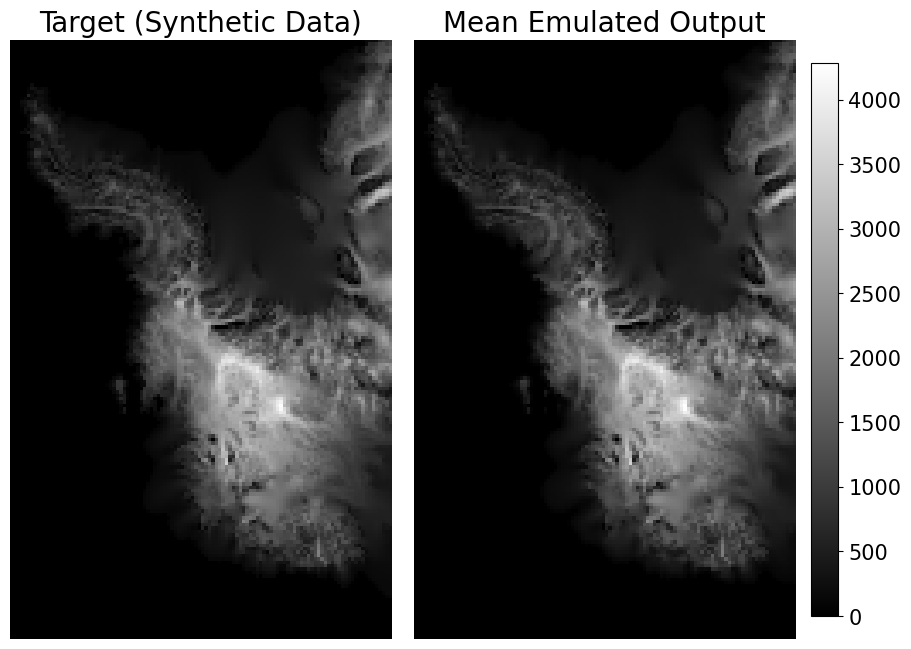}}{%
      \fbox{\parbox{0.55\linewidth}{\centering\vspace{2cm}[figure placeholder: target (left) / emulated mean (right)]\vspace{2cm}}}}
    \caption{An example of the emulation results. The target model run (left) and the pointwise mean ice pattern of the predicted samples (right).}
    \label{fig: Examples for Emulation}
\end{figure}

We evaluate the performance of SC-DS on the full WAIS output: a $176 \times 112$ grid substantially larger than the cropped ASE region in Section~\ref{sec: Comparing with Existing Method}. We first assess the emulator's predictive performance. We randomly hold out 49 model runs from the total of 499 available runs. The remaining 450 model outputs are used to train the emulator, which requires 21.65 hours for 50,000 gradient descent updates. For each of the 49 held-out model runs, we generate 500 emulator outputs and compute the evaluation metrics for the mean emulator output. The overall MAE of the predicted ice thickness maps, computed over locations where the true ice thickness is positive, is approximately 67.32~m. To evaluate the prediction of ice presence and absence, we construct the predicted binary ice maps following the same procedure described in Section~\ref{sec: Comparing with Existing Method}. The mean binary accuracy across the 49 held-out model runs is 96.35\%. These results indicate that the proposed emulator maintains high predictive accuracy even for this much larger domain. Figure~\ref{fig: Examples for Emulation} compares one held-out model run, which closely resembles the observational data, with the corresponding pointwise mean of the emulated ice patterns, showing that the proposed emulator predicts the actual model output accurately. Other model outputs, including those with excessive or sparse ice, are also well captured by our emulator (see Section~D of the Supplementary Material).
 
\begin{figure}[ht]
    \centering
    \IfFileExists{plot/calibration/known_simple.png}{%
      \includegraphics[width=0.48\linewidth]{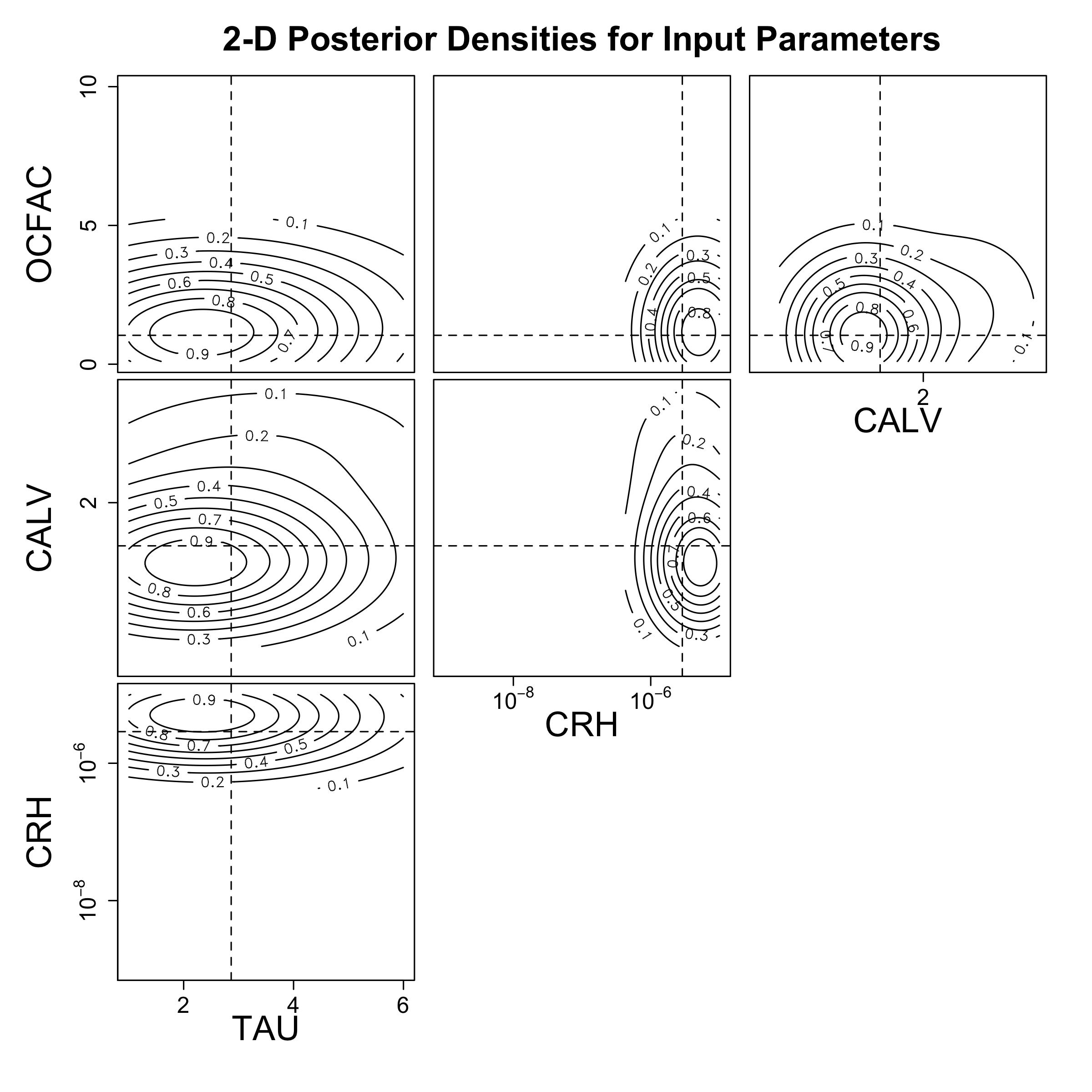}}{%
      \fbox{\parbox{0.55\linewidth}{\centering\vspace{2cm}[figure placeholder: bivariate contour plots, held-out run]\vspace{2cm}}}}
    \caption{The bivariate contour plots for the calibration experiment with a held-out model run. The dashed lines indicate the input parameter settings of the held-out run.}
    \label{fig: Contours with known parameter settings}
\end{figure}

We next diagnose the calibration performance using a held-out model run. As in Section~\ref{sec: Comparing with Existing Method}, this excluded model output is treated as the target synthetic data, and the associated input parameter values serve as the underlying truth for calibration. For this experiment, the emulator is trained on the remaining 498 model runs, and the other computational details are identical to those described in Section~\ref{sec: Comparing with Existing Method}, except for the size and spatial extent of the data considered. The calibration procedure converges after four stages of 1,000 particles each, requiring 18.03 hours of computation. The resulting bivariate marginal density plots for the four selected parameters are shown in Figure~\ref{fig: Contours with known parameter settings}. The pairwise marginal densities are concentrated around the true input parameter values, and compared with the cropped-data experiment in Section~\ref{sec: Comparing with Existing Method}, the densities for all four selected parameters exhibit stronger concentration around the truth. The bivariate marginal density plots for all ten parameters are provided in Section~E of the Supplementary Material.

As in Section~\ref{sec: Comparing with Existing Method}, we further assess the calibration result by generating emulator outputs from the posterior samples. The MAE computed over locations where the target thickness is positive is approximately 93.10~m, and the corresponding binary accuracy is 98.25\%. The pointwise mean ice thickness pattern for this example is provided in Section~D of the Supplementary Material.

\subsection{Results for Observational Data}
\label{sec: Results for Observational Data}
\begin{figure}[ht]
    \centering
    \IfFileExists{plot/calibration/obs_simple.png}{%
      \includegraphics[width=0.48\linewidth]{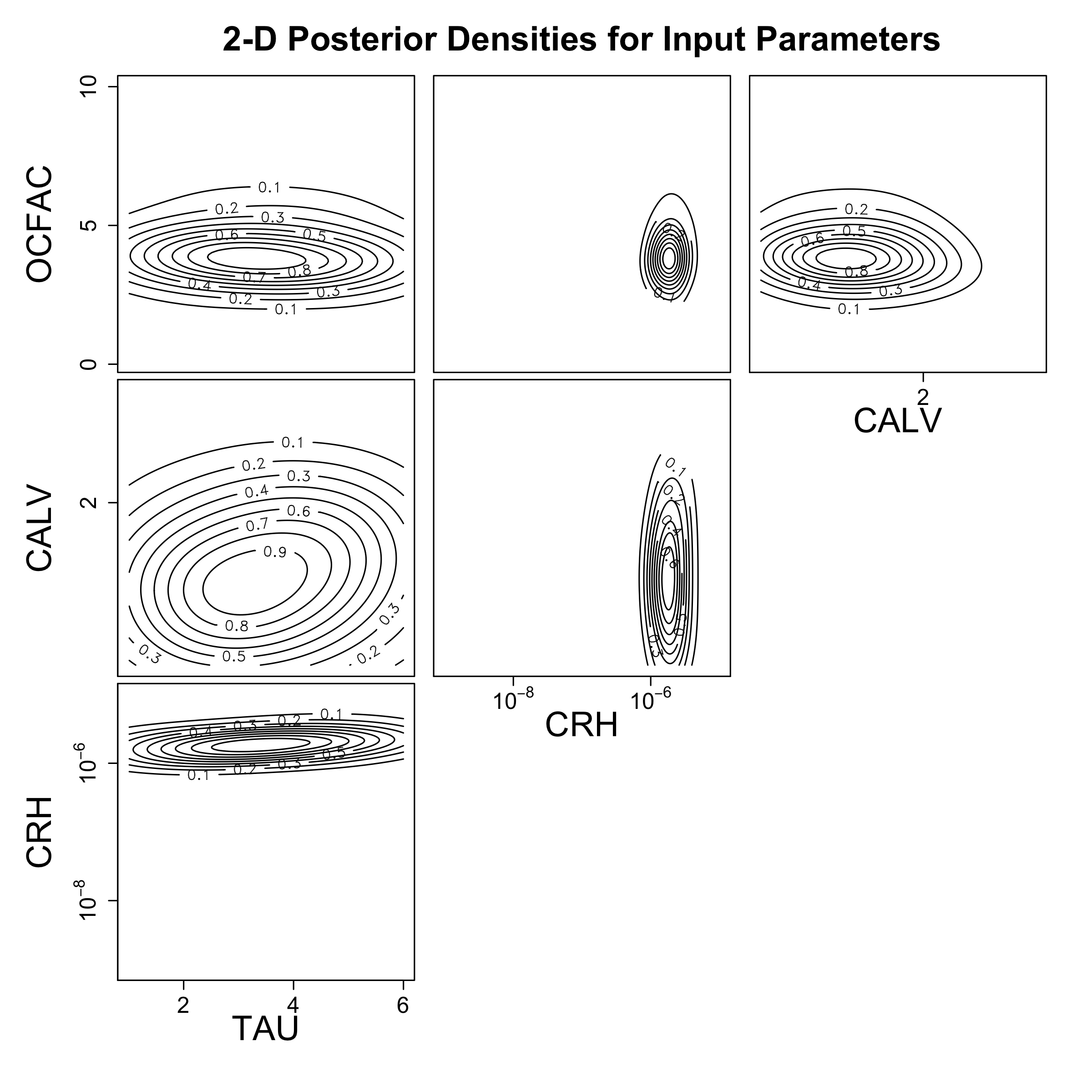}}{%
      \fbox{\parbox{0.55\linewidth}{\centering\vspace{2cm}[figure placeholder: bivariate contours + marginal histograms, Bedmap3]\vspace{2cm}}}}
    \caption{The bivariate contour plots and marginal histograms for the experiment with the Bedmap3 observational data.}
    \label{fig: bfinal_obs4}
\end{figure}

We now apply SC-DS to the observational data (Bedmap3) described in Section~\ref{sec:computer-model-and-data-description}. In this experiment, we use the same emulator as in Section~\ref{sec: Diagnostics of the Performance in Large Scale Model Outputs} and incorporate the discrepancy representation introduced in Section~\ref{sec: Data-model discrepancy}. 
The resulting posterior distributions are shown in Figure~\ref{fig: bfinal_obs4}.
 
Among the four parameters, CRH is the most strongly constrained, with most of its posterior mass concentrated in a relatively narrow range corresponding to larger basal sliding coefficients. This suggests that the observed ice thickness and geometry of the WAIS favor sufficiently fast basal motion across the region. OCFAC is the second most constrained parameter: its posterior density peaks at moderate values (around 4) while ruling out values greater than 6 or less than 2, indicating that the observed WAIS pattern supports oceanic melting of moderate strength. CALV and TAU exhibit notably more diffuse posteriors than the other two, suggesting that the modern observation alone may not provide enough information to sharply constrain them. Nonetheless, their posteriors still concentrate on more plausible regions of the parameter space than a uniform prior would---relatively moderate to slow calving for CALV, and intermediate bedrock rebound times for TAU.
 
\begin{figure}[ht]
    \centering
    \IfFileExists{plot/calibration/Target_Mean_DW.png}{%
      \includegraphics[width=0.8\linewidth]{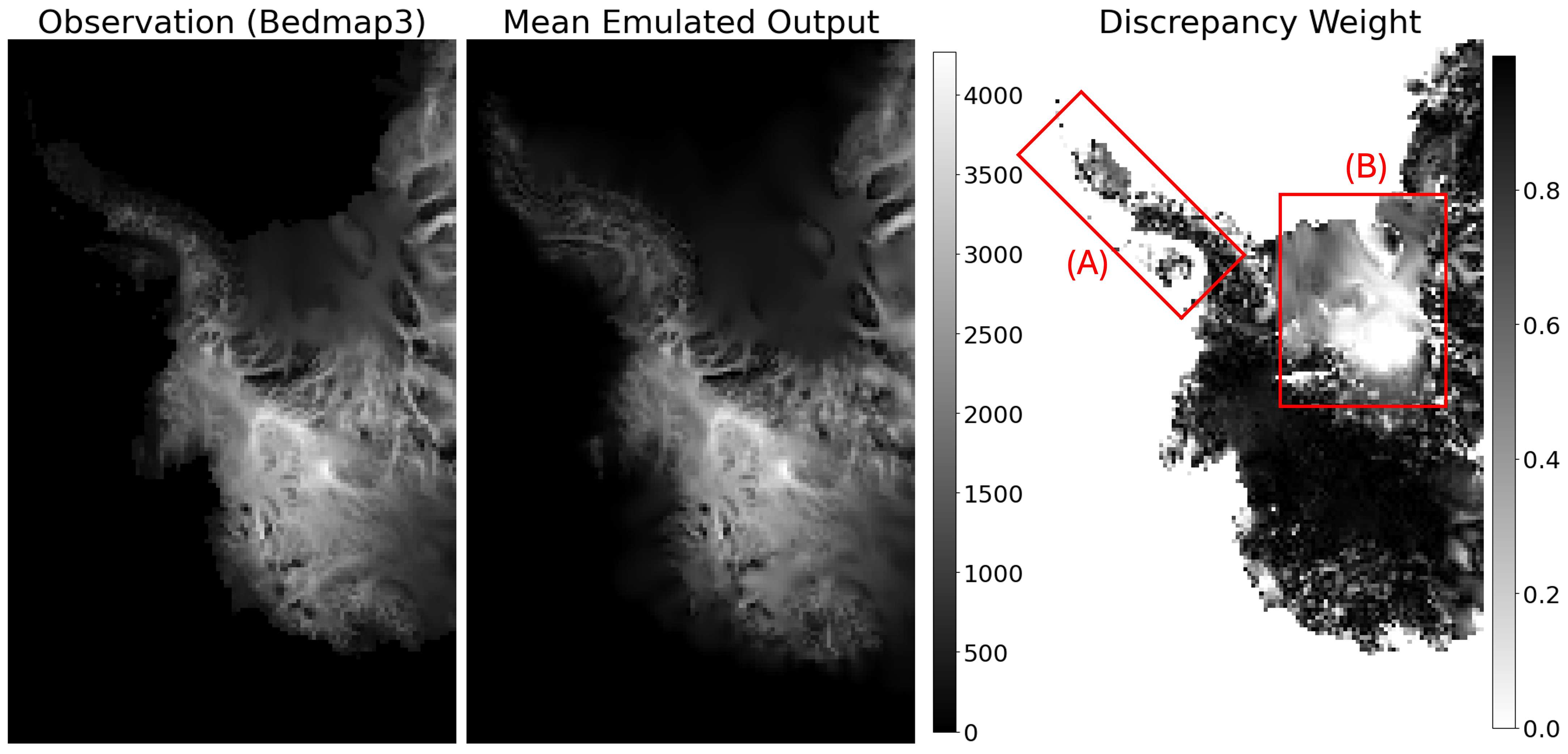}}{%
      \fbox{\parbox{0.85\linewidth}{\centering\vspace{2cm}[figure placeholder: Bedmap3 obs / emulated mean / discrepancy weight map]\vspace{2cm}}}}
    \caption{From left to right: the observations from Bedmap3, the pointwise mean ice pattern of the emulated samples from the posterior, and the discrepancy weight map. Regions with substantial data--model discrepancy, corresponding to low discrepancy weights, are highlighted by the rectangles labeled (A) and (B). Region (A) corresponds to the Antarctic Peninsula (including the Larsen C Ice Shelf region), while region (B) is concentrated around the Filchner--Ronne Ice Shelf.
    }
    \label{fig: predicted observational data}
\end{figure}

We generate 1,000 emulator outputs using parameter values sampled from the posterior distribution to evaluate the calibration performance. Because the calibration incorporates the discrepancy representation computed using the method described in Section \ref{sec: Data-model discrepancy}, we report both the weighted performance, evaluated using the discrepancy-weighted fields, and the unweighted performance, based on the original fields. The weighted results achieve an MAE of approximately 37.75~m at locations where the observed ice thickness is positive and an overall binary accuracy of 99.9\%. When evaluated on the original fields without applying the discrepancy weights, the corresponding values are 82.22~m and 84.15\%, respectively. We attribute the higher weighted accuracy, relative to the simulated example in Section~\ref{sec: Diagnostics of the Performance in Large Scale Model Outputs}, to the down-weighting of high-discrepancy areas, which tend to be the regions where the relationship between the input parameters and the model output is more irregular. In other words, evaluation with the discrepancy weights also down-weights the areas that are more difficult for the emulator to capture. The observational data, the pointwise mean of the emulated ice patterns, and the estimated discrepancy weight map are shown in Figure~\ref{fig: predicted observational data}.
 
Figure~\ref{fig: predicted observational data} shows that the largest discrepancies are concentrated over the Larsen C Ice Shelf on the Antarctic Peninsula and the Filchner--Ronne Ice Shelf. These regions correspond to floating ice shelves, whose geometry is more strongly influenced by ice-shelf dynamics and ice--ocean interactions; they thus involve more complex physical processes than grounded ice. Consequently, they are more difficult to represent accurately using simulation models. By assigning lower weights to these regions, the proposed discrepancy representation reduces their influence during calibration while preserving the information from regions that the model represents well.

\section{Discussion}
\label{sec: Discussion}
In this article, we proposed SC-DS, a neural network-based likelihood-free framework for the emulation and calibration of computer models. The proposed methodology is designed for high-dimensional semi-continuous spatial data, for which existing calibration approaches are often computationally infeasible. Although our primary motivation is calibrating an Antarctic ice-sheet model, the proposed framework is broadly applicable, with only minor modifications, to calibration problems involving high-dimensional spatial outputs.
 
For emulation, we employed a conditional diffusion model to approximate the PSU3D-ICE model over a multidimensional parameter space. By exploiting the latent representation of the diffusion model, the proposed architecture jointly models the zero-inflated and continuous components of the spatial output within a unified framework. We further introduced complementary global and local conditioning mechanisms that let the emulator represent both domain-wide and spatially localized effects of the input parameters. For calibration, we developed a likelihood-free framework that replaces the deterministic acceptance rule of conventional ABC with a probabilistic acceptance rule learned by a Siamese network. The resulting similarity measure is incorporated into a sequential calibration procedure together with a discrepancy representation, yielding an approximate posterior distribution without requiring either an explicit likelihood function or numerical integration over the data distribution.
 
The numerical studies demonstrate that SC-DS produces posterior distributions comparable to those of the current state-of-the-art GP-based approach on the reduced spatial domain while substantially reducing computational cost. More importantly, the proposed method remains computationally feasible for the full WAIS domain, where likelihood-based approaches become prohibitively expensive. The simulated example with the full WAIS domain confirms that SC-DS can accurately recover the true parameter values, and the calibration results using Bedmap3 observations are physically interpretable and consistent with the current understanding of WAIS dynamics.
 
Several directions remain for future research. Because the overall calibration performance depends strongly on the fidelity of the diffusion emulator, further improvements may be achieved through more advanced diffusion architectures and training strategies that better reflect the underlying physics of the computer model. In addition, the current framework focuses on calibrating the present-day ice-sheet geometry from a single spatial field. Extending the proposed methodology to incorporate time-evolving observations, such as satellite altimetry or ice-velocity measurements, would enable calibration based on both spatial and temporal information and provide additional constraints on parameters governing the transient evolution of the ice sheet.


\section*{Supplementary Material}
The online supplementary material contains additional model details and supplementary figures.


\section*{Acknowledgement}
We would like to thank the Editor, the Associate Editor, and the anonymous reviewers for their careful reading and valuable comments.


\section*{Funding}
Kanghyun Wi and Jaewoo Park were supported by the National Research Foundation of Korea (RS-2025-00513129) and ICAN (ICT Challenge and Advanced Network of HRD) support program (RS-2023-00259934) supervised by the IITP (Institute for Information \& Communication Technology Planning \& Evaluation). Won Chang was supported by the National Research Foundation of Korea (NRF) grant funded by the Ministry of Science and ICT (MSIT) (RS-2025-00523567), the Global-LAMP Program of the National Research Foundation of Korea (NRF) grant funded by the Ministry of Education (RS-2023-00301976), and New Faculty Startup Fund from Seoul National University (326-20240027).

\section*{Data Availability Statement}\label{data-availability-statement}
The source code and data are available at {\url{https://github.com/khwi97/E-and-C-PSU-3D-DFM2}}.


\section*{Disclosure statement}
The authors report there are no competing interests to declare. ChatGPT and Claude were used to support language editing throughout the article.

\clearpage
\appendix

\begin{center}
\title{\LARGE\bf Supplementary Material for ``Scalable, Likelihood-Free Calibration of Ice-Sheet Models with Deep Diffusion Emulators and Feature Matching''}\\~\\
  
\author{\Large{Kanghyun Wi, Jaewoo Park, Saumya Bhatnagar, and Won Chang}}
\end{center}
\maketitle

The supplementary material contains additional model details and supplementary figures.

\section{Table of Input Parameters for PSU-3D model}

\begin{table}[htbp]
    \centering
    \resizebox{\textwidth}{!}{
    \begin{tabular}{cc}
    \toprule
        Parameters & Definition \\
    \midrule
        OCFAC & Ocean sub-ice shelf melt factor, 0.1 to 10, non-dimensional\\
        CALV & Calving factor, 0.316 to 3.16, non-dimensional\\
        CRH & Basal sliding coefficient for the modern ocean region, $10^{-9}$ to $10^{-5}$, m $\mathrm{yr^{-1}Pa^{-2}}$\\
        CRHASE & Extra multiplicative factor for CRH only in the inshore of Amundsen Sea Embayment (ASE), $10^{-3}$ to $1$, non-dimensional \\
        TAU & Asthenospheric relaxation $e$-folding time scale, 1 to 6 kyr\\
        LITH & Lithospheric stiffness, $10^{23}$ to $10^{25}$, N m\\
        GEO & Geothermal heat flux, 50 to 90, mW $\mathrm{m^{-2}}$\\
        SUBPIN & Sub-grid ice-shelf pinning factor, 0 to 4, non-dimensional\\
        USCH & Grounding-line flux factor, 0.5 to 5, non-dimensional\\
        LAPSE & Atmospheric lapse rate, $-0.005$ to $-0.010$, $^{\circ}\mathrm{Cm^{-1}}$\\
    \bottomrule
    \end{tabular}
    }
    \caption{All of 10 calibrated parameters with definition, ranges, and units. The parameters called factors are non-dimensional. The ranges are scaled to $[0,1]$ in our experiments. More descriptions of parameters with the equations can be found in \citep{PollardandDeConto2012, PollardEtAl2015}.}
    \label{tab: Description of parameters}
\end{table}

\section{Deep Neural Diffusion Model Architectures Used in the Emulation Steps}
\begin{figure}
    \centering
    \includegraphics[width=\linewidth]{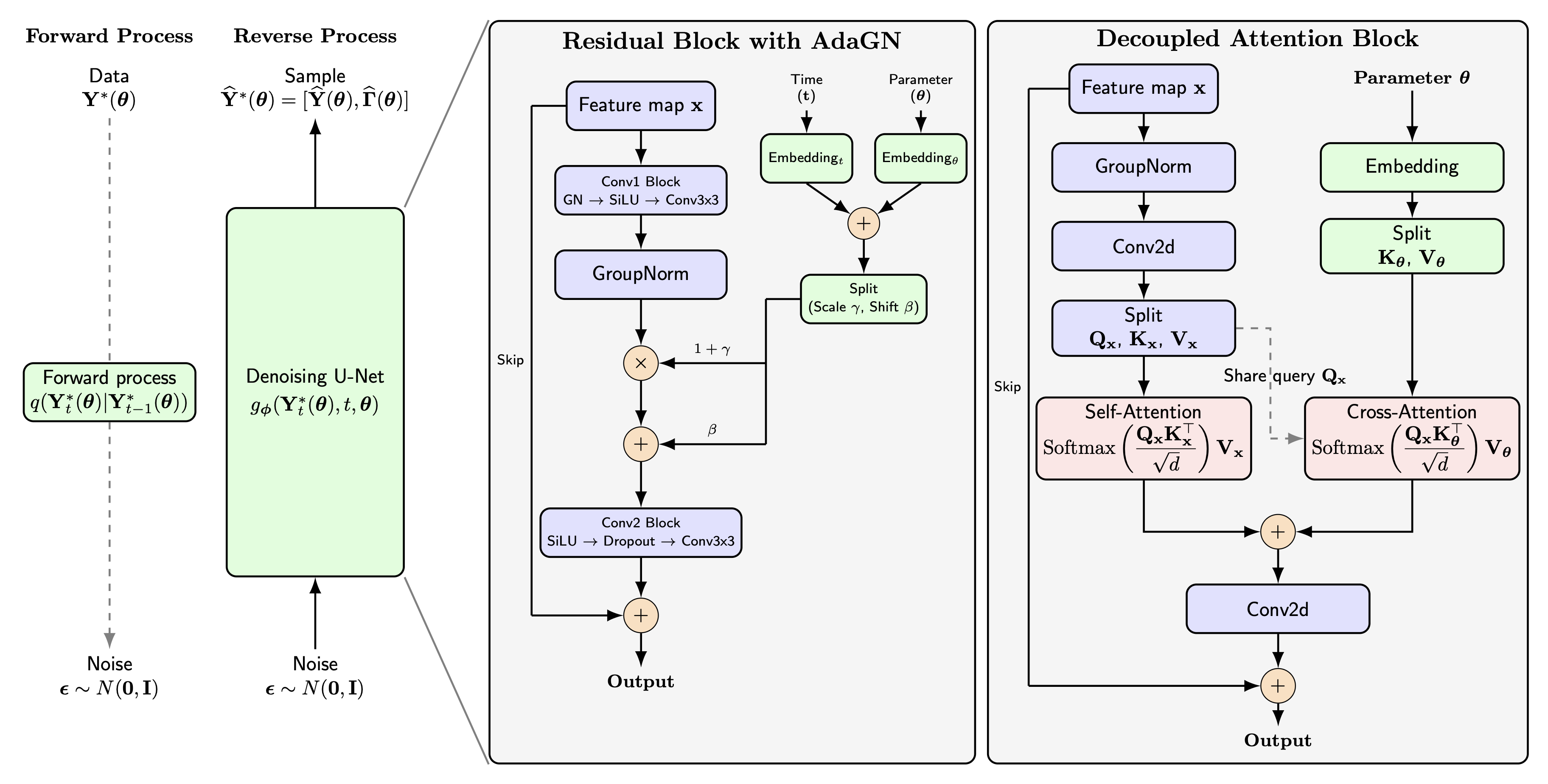}
    \caption{The diffusion model architecture for the emulation step. The denoising U-net is defined as a composition of residual blocks and attention blocks within the encoder and decoder architectures, as described in Section 3.}
    \label{fig: DiffusionArchitecture}
\end{figure}

\section{Details of the Siamese Network in Section 4.2}
In this subsection, we describe the architecture of the CNN that composes the Siamese network used in the calibration stage (Section 4.2). Each weight-sharing convolutional neural network consists of $L$ convolution layers followed by $R$ fully connected layers. We define the $\ell$-th convolution layer of a CNN as
\begin{equation}
\label{eq: convolutional layer}
    \boldsymbol{\Lambda}^{(\ell)}= h^{(\ell)} \left[ \psi_1^{(\ell)} \left( \mathbf{K}^{(\ell)}\ast \boldsymbol{\Lambda}^{(\ell-1)} + \mathbf{B}^{(\ell)}\right)\right],
\qquad \ell=1,\dots,L,
\end{equation}
where $\ast$ is a convolution operator, $\boldsymbol{\Lambda}^{(\ell)}\in \mathbb R^{C^{(\ell)}\times H^{(\ell)}\times W^{(\ell)}}$ is the feature map of the $\ell$-th convolution layer, and $\mathbf{K}^{(\ell)}\in \mathbb{R}^{C^{(\ell)} \times C^{(\ell-1)} \times k^{(\ell)} \times k^{(\ell)}}$ and $\mathbf{B}^{(\ell)} \in \mathbb R^{C^{(\ell)} \times H^{(\ell - 1)} \times W^{(\ell - 1)}}$ are the convolution kernel and bias for the $\ell$-th layer, respectively. Furthermore, $\psi_1^{(\ell)}(\cdot)$ is an activation function, and $h^{(\ell)}(\cdot)$ denotes a spatial down-sampling operation such as max pooling or average pooling. The initial input is $\boldsymbol{\Lambda}^{(0)}=\{\mathbf{x}\}$, where $\mathbf{x}$ denotes a generic $n$-dimensional input field (either $\mathbf{U}$ or $\mathbf{V}$) and $\{\cdot\}$ is a reshaping operator that rearranges the $n$-dimensional vector onto the original grid space $\mathcal{S}$. 

To identify informative regions in the target ice pattern, we incorporate a spatial attention module inspired by \cite{woo2018cbam} into the convolutional layers. Suppose that the attention module is applied to the feature map $\boldsymbol{\Lambda}^{(\ell)}$. Let $\boldsymbol{\Lambda}_{\mathrm{avg}}^{(\ell)},
\boldsymbol{\Lambda}_{\mathrm{max}}^{(\ell)}
\in \mathbb R^{1\times H^{(\ell)}\times W^{(\ell)}}$
be the outputs of average pooling and max pooling applied to $\boldsymbol{\Lambda}^{(\ell)}$ over the channel dimension $C^{(\ell)}$, respectively, and let $\boldsymbol{\Lambda}^{(\ell)}_S\in \mathbb R^{2\times H^{(\ell)}\times W^{(\ell)}}$ be their concatenation. The attention map is computed as
\begin{equation}
    \nonumber
    \mathbf M^{(\ell)} = \sigma\left( \mathbf{K}_S \ast \boldsymbol{\Lambda}^{(\ell)}_S \right),
\end{equation}
where $\sigma(\cdot)$ is an element-wise sigmoid function and $\mathbf{K}_S\in \mathbb R^{1\times 2\times k_S\times k_S}$ is the convolution kernel. Here $\mathbf{K}_S$ follows the same four-dimensional convention as $\mathbf{K}^{(\ell)}$ in main article, whose leading two dimensions denote the numbers of output and input channels; it maps the two-channel input $\boldsymbol{\Lambda}^{(\ell)}_S$ to a single-channel attention map, so the output- and input-channel dimensions are $1$ and $2$, respectively. The resulting attention map $\mathbf M^{(\ell)}\in\mathbb R^{1\times H^{(\ell)} \times W^{(\ell)}}$ likewise retains an explicit singleton channel dimension so that it can be broadcast (i.e., multiplied element-wise) across all $C^{(\ell)}$ channels of $\boldsymbol{\Lambda}^{(\ell)}$, such that
\begin{equation}
    \label{eq: spatial attention}
    \widetilde{\boldsymbol{\Lambda}}^{(\ell)}_c= \mathbf M^{(\ell)} \odot \boldsymbol{\Lambda}^{(\ell)}_c,\quad c=1,\dots, C^{(\ell)},
\end{equation}
where $\odot$ is the element-wise product operator and the subscript $c$ indexes the channel (first) dimension of $\boldsymbol{\Lambda}^{(\ell)}$ and $\widetilde{\boldsymbol{\Lambda}}^{(\ell)}$, so that ${\boldsymbol{\Lambda}}^{(\ell)}_c, \widetilde{\boldsymbol{\Lambda}}^{(\ell)}_c\in \mathbb R^{1\times H^{(\ell)}\times W^{(\ell)}}$. In our implementation, the attention in \eqref{eq: spatial attention} is applied after the second convolution layer, and $\widetilde{\boldsymbol{\Lambda}}^{(\ell)}$ is then treated as the input feature map of the next convolution layer. Because $\mathbf{K}_S$ is a small, spatially shared kernel, it does not encode fixed positions but instead learns to produce an \emph{input-dependent} attention map $\mathbf M^{(\ell)}$ that up-weights locations most informative for distinguishing parameter settings and suppresses regions that are uninformative or noisy. As the same $\mathbf{K}_S$ is used in both branches, the corresponding salient regions are emphasized consistently in the two input fields before they are compared, which sharpens the learned similarity measure.

The extracted features are then passed through $R$ fully connected layers. Define $\boldsymbol{\lambda}^{(0)}$ as the $q$-dimensional vectorized output of the $L$-th convolution layer $\boldsymbol{\Lambda}^{(L)}$, where $q=C^{(L)}\times H^{(L)}\times W^{(L)}$. The $r$-th output of the fully connected layers, denoted $\boldsymbol{\lambda}^{(r)}$, is obtained as
\begin{equation}
    \label{eq: fully connected layer}
    \boldsymbol{\lambda}^{(r)} = \psi_2^{(r)}(\mathbf{W}^{(r)}\boldsymbol{\lambda}^{(r-1)} + \mathbf{b}^{(r)}), \quad r=1,\dots,R,
\end{equation}
where $\psi_2^{(r)}(\cdot)$ is an activation function and $\mathbf{W}^{(r)}\in \mathbb R^{q^{(r)}\times q^{(r-1)}}$ and $\mathbf{b}^{(r)}\in\mathbb R^{q^{(r)}}$ are the weight matrix and bias vector for the $r$-th fully connected layer, respectively. Applying this shared branch to the two input fields yields the feature vectors $\boldsymbol{\lambda}_{\mathbf{U}}$ and $\boldsymbol{\lambda}_{\mathbf{V}}$.

\section{Comparison of the Pointwise Mean Ice Thickness and Target Data}
\begin{figure}
    \centering
    \includegraphics[width=\linewidth]{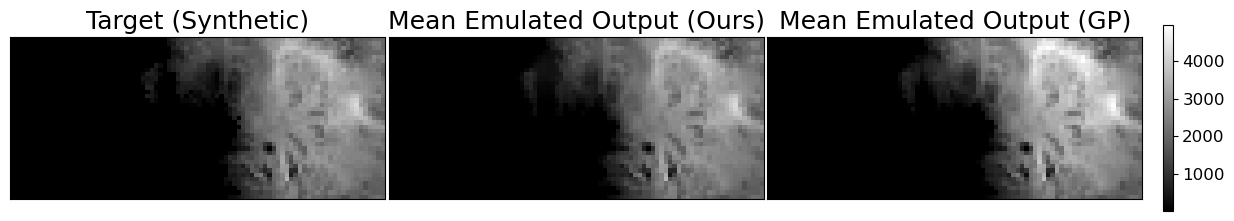}
    \caption{From left to right: the target synthetic data, the pointwise mean ice thickness pattern obtained from the proposed method, and the pointwise mean ice thickness pattern obtained from the existing method of \cite{Chang2022}. The mean predicted patterns are computed by emulating 1,000 parameter samples from each calibration method discussed in Section 5.1.}
    \label{fig: ComparingPWGP}
\end{figure}

\begin{figure}
    \centering
    \includegraphics[width=0.8\linewidth]{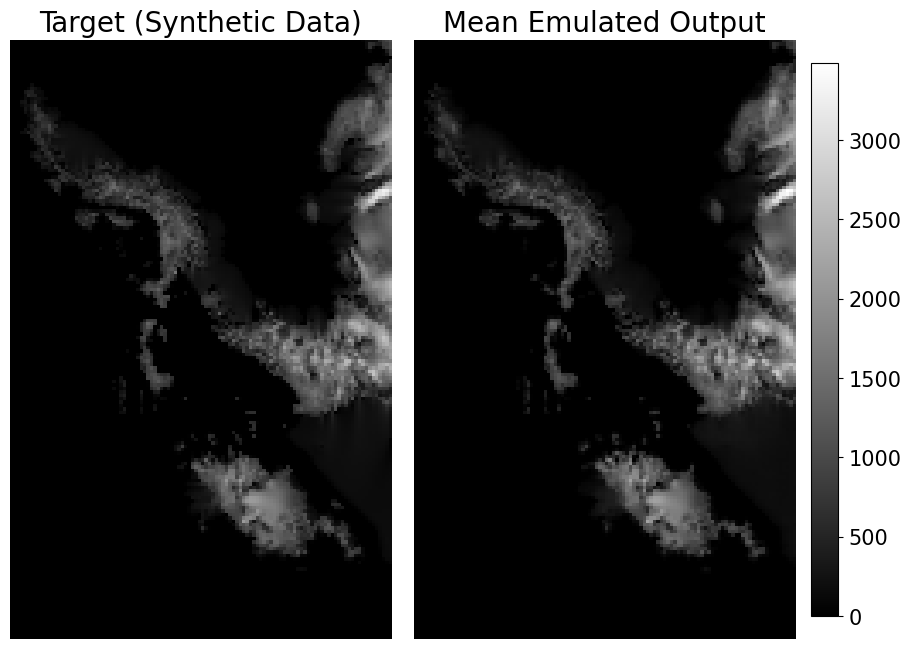}
    \includegraphics[width=0.8\linewidth]{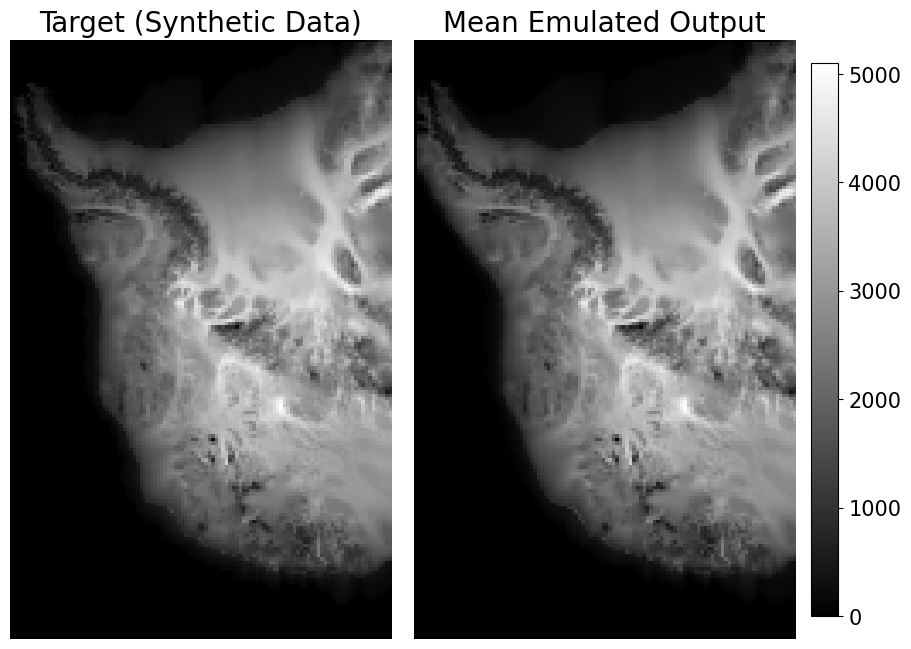}
    \caption{Examples of the emulation step described in Section 5.2. (Top): The target model run with sparse ice (left) and the pointwise mean ice pattern of the predicted samples (right). (Bottom): The target model run with excessive ice (left) and the pointwise mean ice pattern of the predicted samples (right). We generate 500 emulator outputs for the corresponding pointwise mean ice maps.}
\end{figure}

\begin{figure}
    \centering
    \includegraphics[width=\linewidth]{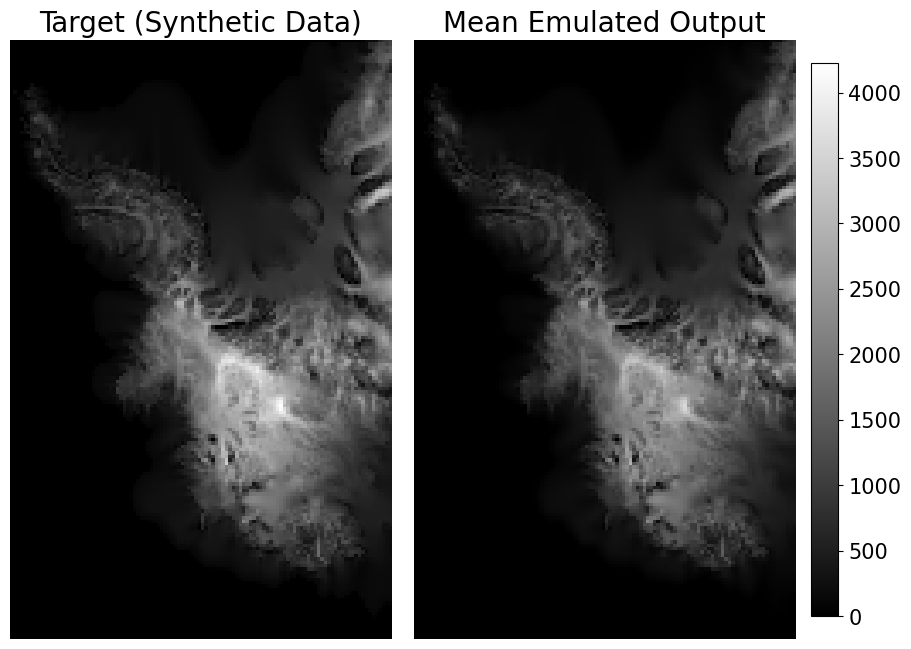}
    \caption{The target computer model run (left), and the pointwise mean ice thickness pattern obtained using the proposed emulator and calibration method (right). The mean predicted patterns are computed by emulating 1,000 parameter samples from the posterior discussed in Section 5.2.}
\end{figure}

\section{Posterior Densities for All Ten Parameters}
\begin{figure}
    \centering
    \includegraphics[width=\linewidth]{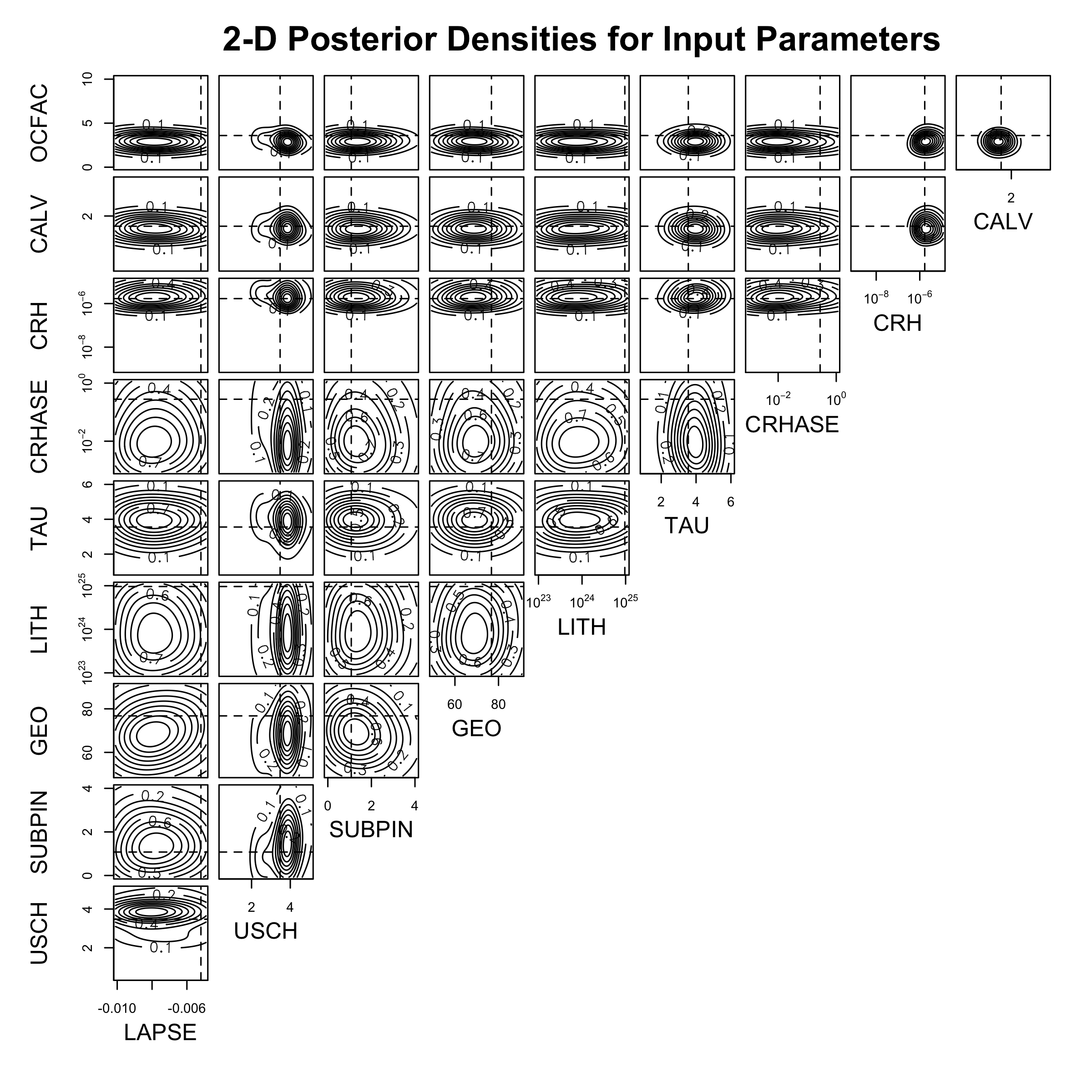}
    \caption{Two-dimensional contours of the input parameters $\boldsymbol{\theta}$ obtained using the MCMC methods proposed in \cite{Chang2022}, based on the synthetic data example described in Section 5.1. The dashed lines represent the true input parameters. }
    \label{fig: 10parameterGP}
\end{figure}

\begin{figure}
    \centering
    \includegraphics[width=\linewidth]{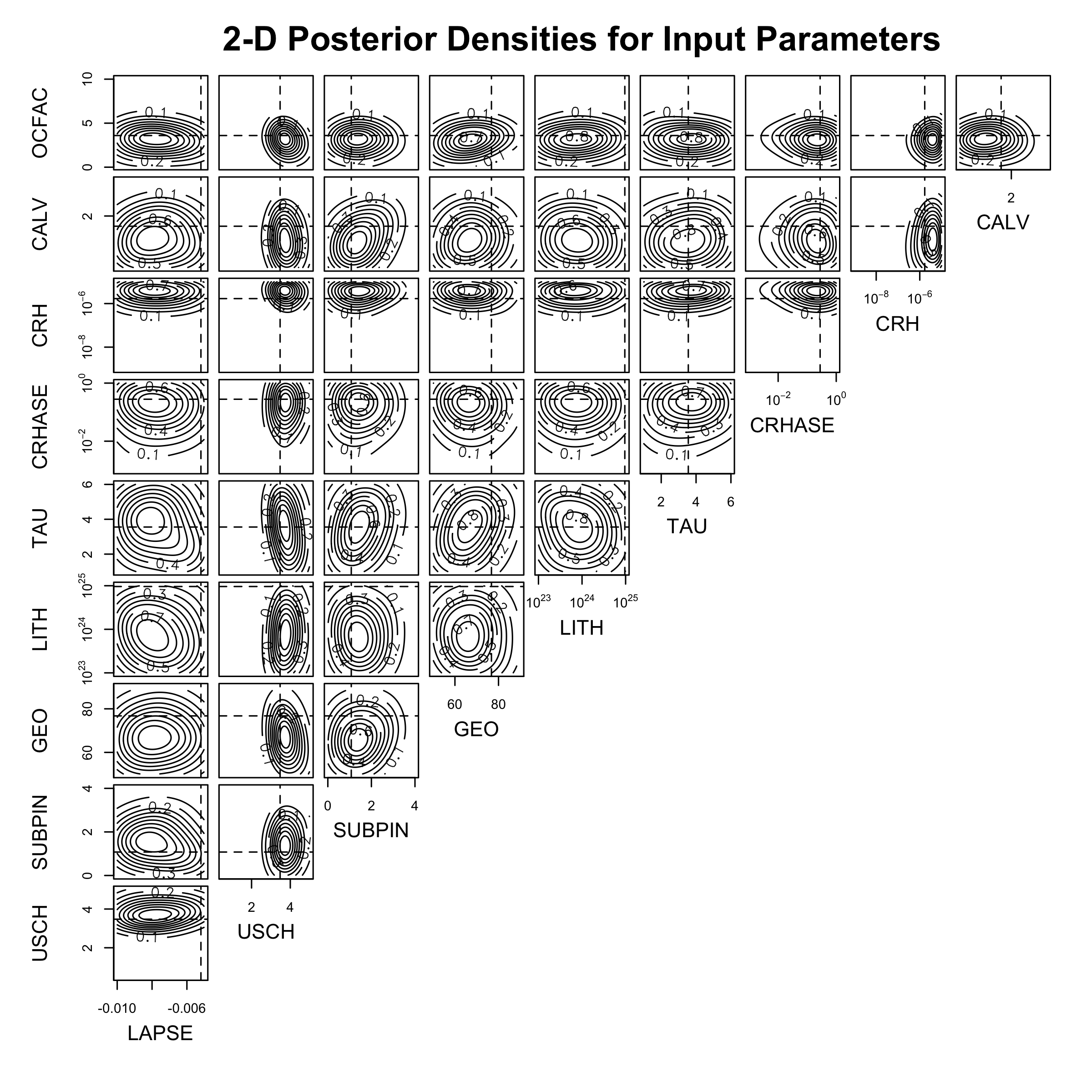}
    \caption{Two-dimensional contours of the input parameters $\boldsymbol{\theta}$ obtained using our proposed methods, based on the synthetic data example described in Section 5.1. The dashed lines represent the true input parameters.}
    \label{fig: 10parameterOur}
\end{figure}

\begin{figure}
    \centering
    \includegraphics[width=\linewidth]{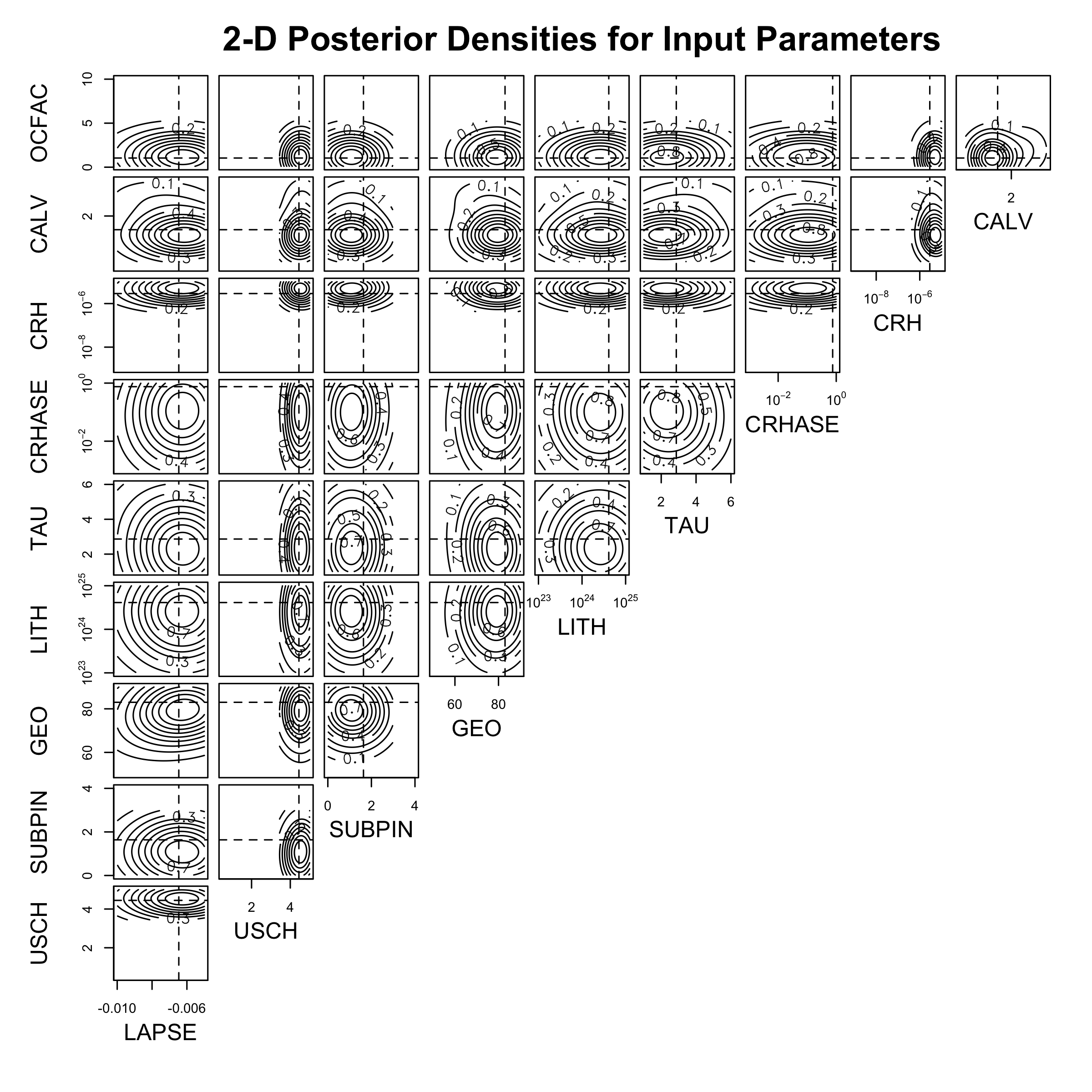}
    \caption{Two-dimensional contours of the input parameters $\boldsymbol{\theta}$ obtained using our proposed methods, based on the computer model output described in Section 5.2. The dashed lines represent the true input parameters.}
    \label{fig: 10parameterOur52}
\end{figure}

\begin{figure}
    \centering
    \includegraphics[width=\linewidth]{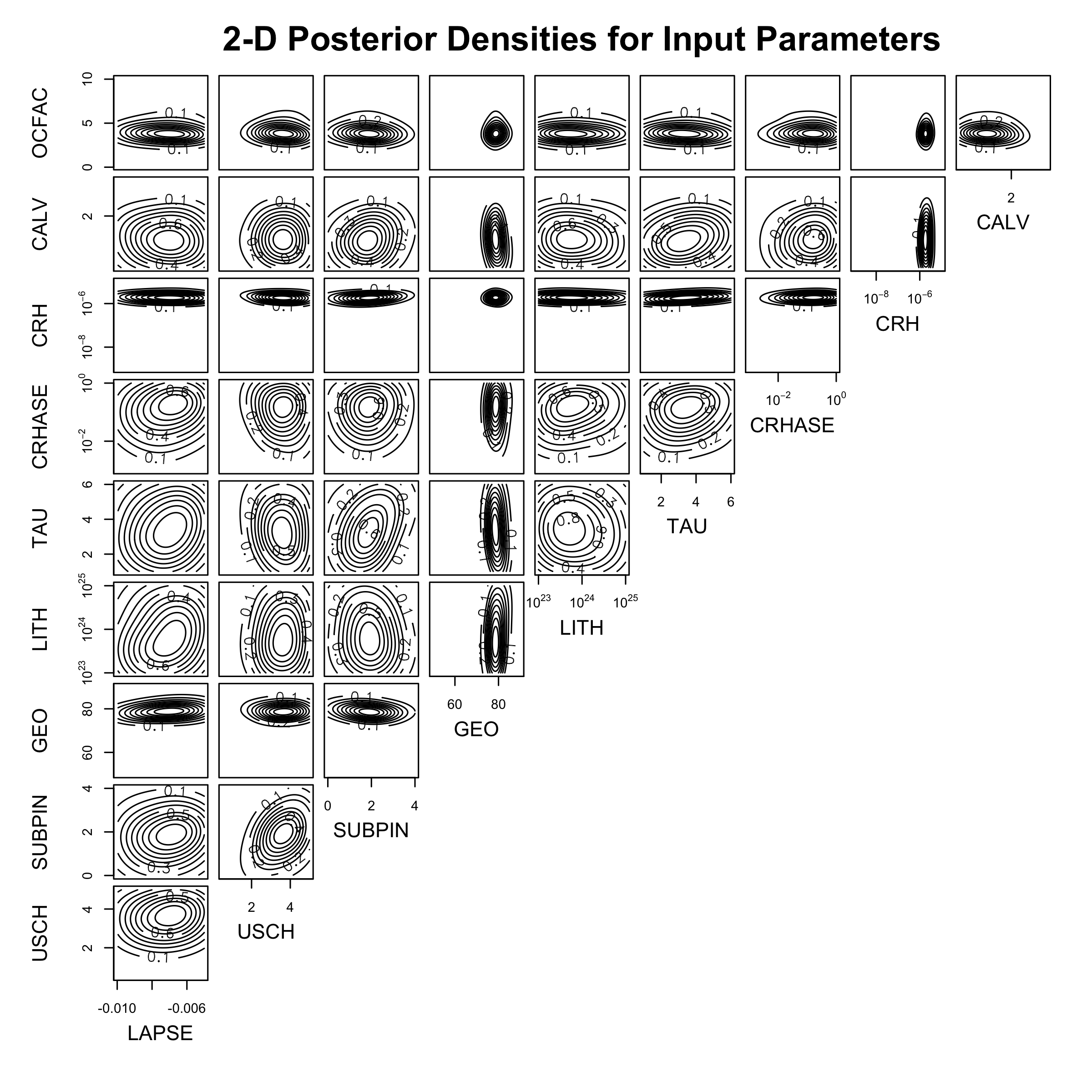}
    \caption{Two-dimensional contours of the input parameters $\boldsymbol{\theta}$ obtained using our proposed methods, based on the observational data (Bedmap 3) described in Section 5.3. For the detailed description of the observational data, please see Section 2 and \cite{bedmap3}.}
    \label{fig: 10parameterOur53}
\end{figure}
\bibliography{ref}

\end{document}